\documentclass[%
 reprint,
superscriptaddress,
 amsmath,amssymb,
 aps,
]{revtex4-2}

\usepackage{amsmath}
\usepackage{slashed}
\usepackage{mathtools}
\usepackage{lipsum}
\usepackage[hidelinks,colorlinks=true,linkcolor=blue,citecolor=blue,urlcolor=blue]{hyperref}
\usepackage{wasysym}
\usepackage{multirow}
\usepackage[T1]{fontenc}

\begin{document}

\title{Relativistic Mean Field Model parameterizations in the light of GW170817, GW190814, and 
PSR J0740 + 6620}
\author{Virender Thakur}
 \email{virenthakur2154@gmail.com}
 \affiliation{Department of Physics, Himachal Pradesh University, Shimla-171005, India}
 \author{Raj Kumar}
  \email{raj.phy@gmail.com}%
\affiliation{Department of Physics, Himachal Pradesh University, Shimla-171005, India}
\author{Pankaj Kumar}
 \affiliation{Department of Physics, Himachal Pradesh University, Shimla-171005, India}
\author{Vikesh Kumar}
\affiliation{Department of Physics, Himachal Pradesh University, Shimla-171005, India}
\author{B.K. Agrawal}
\affiliation{Saha Institute of Nuclear Physics, 1/AF Bidhannagar, Kolkata 700064, India}
\author{Shashi K. Dhiman}
 \email{shashi.dhiman@gmail.com}
\affiliation{Department of Physics, Himachal Pradesh University, Shimla-171005, India}
\affiliation{School of Applied Sciences, Himachal Pradesh Technical University, Hamirpur-177001, India}

\begin{abstract}
Three parameterizations  DOPS1, DOPS2, and DOPS3 (named after  the Department of Physics Shimla)  of the Relativistic Mean Field (RMF) model have been proposed with the inclusion of all possible self and mixed interactions between the scalar-isoscalar ($\sigma$), vector-isoscalar ($\omega$) and vector-isovector ($\rho$) mesons up to quartic order. The generated parameter sets are in harmony with the finite and bulk nuclear matter properties.
 A set of Equations of State (EOSs) composed of pure hadronic (nucleonic) matter  and nucleonic with quark matter (hybrid EOSs) for superdense hadron-quark
matter in $\beta$-equilibrium is obtained. The quark matter phase is calculated  by using the three-flavor Nambu-Jona-Lasinio (NJL) model.  The maximum mass of a non-rotating neutron star with DOPS1 parameterization is found to be around 2.6 M$\odot$ for the pure nucleonic matter which satisfies the recent gravitational wave analysis of  GW190814  [\href{https://iopscience.iop.org/article/10.3847/2041-8213/ab960f}{Abbott et al., Astrophys. J. Lett. {\bf{896}}, L44 (2020)}]  with
possible maximum mass constraint indicating  that the secondary component of GW190814 could be a non-rotating heaviest neutron star composed of pure nucleonic matter. EOSs computed with the DOPS2 and DOPS3 parameterizations satisfy the X-Ray observational data [\href{https://iopscience.iop.org/article/10.1088/0004-637X/722/1/33}{Steiner et al., Astrophys. J. {\bf{722}}, 33 (2010)}] and the recent  observations of GW170817  maximum mass constraint of a stable non-rotating neutron star in the range 2.01 $\pm$ 0.04 - 2.16 $\pm$ 0.03 M$\odot$    [\href{https://iopscience.iop.org/article/10.3847/2041-8213/aaa401/meta}{Rezzolla et. al., Astrophys. J. Lett.  {\bf{852}}, L25 (2018)}] and also in good agreement with constraints on mass and radius measurement for PSR J0740+6620 (NICER)   [\href{https://iopscience.iop.org/article/10.3847/2041-8213/ac0a81/meta}{Riley et al., Astrophys. J. Lett.  {\bf{918}}, Riley et al.,  L27 (2021)},\href{https://iopscience.iop.org/article/10.3847/2041-8213/ac089b}{Miller et al., Astrophys. J. Lett.  {\bf{918}}, L28 (2021)}]. The hybrid EOSs obtained with the NJL model also satisfy astrophysical constraints on the maximum mass of a neutron star from PSR J1614-2230   [\href{https://www.nature.com/articles/nature09466}{Demorest et al., Nature {\bf{467}}, 1081 (2010)}].
We also present the  results for dimensionless tidal deformability,
${\Lambda}$  which are consistent
with the waveform models analysis of GW170817. 
\end{abstract}
\maketitle


\section{Introduction}
\label{i}

The knowledge of neutron star properties is necessary to probe the high density behavior of the equations of state (EOSs) for the baryonic matter in the beta equilibrium. Neutron  stars  are  the  densest  manifestations of massive objects in the observable universe and  sound  knowledge  of     
 EOSs of dense matter is required to understand the properties of neutron stars.
 The precise gravitational mass and radius measurements of the neutron stars are 
 effective ways to constrain the EOSs of high dense matter in its interiors. 
The  mass measurement of MSP J0740+6620  \citep{Cromartie2020} with 
2.14$^{+0.10}_{-0.09}M_{\odot}$,  is likely to be the most massive neutron star yet observed. Recently, the simultaneous measurements of gravitational mass M and equatorial circumferential radius R$_{eq}$ 
of PSR J0030+0451 from NICER data by Miller et al.  \citep{Miller2019} and Riley et al.  \citep{Riley2019} 
by using independent methods to actual map of the hot region of pulsar, have inferred [M = 1.44$^{+0.15}_{-0.14}$M$_\odot$, R$_{eq} = 13.02^{+1.24}_{-1.06}$km]  and  [M = 1.34$^{+0.15}_{-0.16}$M$_\odot$, R$_{eq} = 12.71^{+1.14}_{-1.19}$km], respectively.   
\par Theoretically, the investigations of the observed masses and  radii of Compact Stars (CS) reveals the particle composition and phase transition of dense nuclear matter at high densities. Several attempts 
  \citep{Dhiman2007,Ozel2016,Oertel2017,Annala2018} have been made to construct the EOSs comprising of nucleons, hyperons and quarks under the constraint of global $\beta$-equilibrium.
The inclusion of hyperons and/or quarks in EOSs softens the high density behavior, leading to the reduction of maximum gravitational masses of CS. 
Recently, there are many EOS models that include hyperons as well as  quark matter   \citep{Oertel2017,Alvarez2019} and  
maximum gravitational mass calculated from them is  compatible with $\approx $ 2M$_{\odot}$. 
\par
The theory of strong interactions, quantum chromodynamics (QCD), and ultra-relativistic heavy-ion collisions  predict that at high  densities, the hadronic  matter may undergo a deconfinement phase consisting of quarks and gluons. Therefore, recently, it is an open question whether the inner core of compact stars (CS) consists of quark matter 
   \citep{Witten1984,Farhi1984,Haensel1986,Alcock1986,Lattimer2004}. However, this has been suggested currently that the dense nuclear matter in the interior of stable compact stars with maximum gravitational masses M$\approx$ 2.0M$_{\odot}$ may exhibit evidence for the presence of quark matter cores   \citep{Annala2018}.    
 Therefore, the hybrid stars phenomenology offers a unique tool to address the 
 challenge of understanding the phase transition in  dense quantum chromodynamics. The nuclear theory studies   \citep{Haensel2007,Lattimer2014,Baym2018} 
are mainly focusing on understanding the  dense matter of compact stars (CS). 
The recent observations with LIGO and Virgo of GW170817 event   \citep{Abbott2018,Abbott2019} of Binary Neutron Stars merger
and the discovery of CS with masses around 2$M_\odot$   \citep{Demorest2010,
Antoniadis2013,Arzoumanian2018,Miller2019,Riley2019, Raaijmakers2019} have intensified the interest in these 
astonishing  objects. The analysis of GW170817 has demonstrated the potential
of gravitational wave (GW) observations to yield new information relating to 
the limits on CS tidal deformability. In addition to these astrophysical 
observations   \citep{Champion2008,Abdo2010,Guillemot2012,Fonseca2016,Ozel2016},
the measurements of rotation frequencies of the pulsar can be employed to 
constraints the particle composition and behavior of EOSs of the dense nuclear matter. However, the direct measurement
of radius and quark matter interior core of CS is still a great challenge from astrophysical interests.
In many papers, the properties of cold quark matter have been studied in terms of the phenomenological
MIT quark bag model and EOSs at zero temperature have been obtained; these are the basis of
calculations of the characteristics of hybrid hadron-quark stars, as well as of strange quark stars \citep{9,10,11,15,16}.
The NJL model \citep{19,20} has recently often been used to describe quark matter; it was originally proposed for
explaining the origin of the nucleon mass taking the spontaneous violation of chiral symmetry into account and
was later reformulated for the description of quark matter \citep{22,21}. This model successfully reproduces many
features of QCD \citep{24,23}. Combining different modifications of the NJL quark model with different models for
describing hadron matter, several authors have constructed hybrid EOSs of cold matter and used these to
study the properties of neutron stars containing quark matter \citep{28,26,29}.
\par The quark matter phase of 
EOSs have been treated by employing phenomenological
models with some basic features of QCD, such as, 
the MIT bag models   \citep{Alford2005,Agrawal2009,Zhou2018} with a bag constant and
appropriate perturbative QCD corrections and 
Nambu-Jona-Lasinio with chiral symmetry and its breaking   \citep{Menezes2006}, Non-local 
chiral quark model   \citep{Blaschke2007} and  constant speed of sound model   \citep{Alford2013}.

The motivation of the present work is  to compute a set of EOSs where the hadronic
phase has been calculated within the framework of energy density functionals based
on the RMF theory   \citep{Dhiman2007} and, the quark matter phase
of EOS is computed by using three flavor Nambu-Jona-Lasinio (NJL) model with scalar-isovector and vector-isovector couplings. A  plausible set of EOSs for hadron-quark
matter is employed to study  the structural properties of non-rotating neutron stars which satisfies the astrophysical constraints of GW170817, GW190814, PSR J0740+6620, and other available observational data.
\par
The RMF model used in the present work  includes all possible self and mixed interaction terms for the $\sigma$, $\omega$, and $\rho$ mesons. The $\omega$ meson self-coupling term enables one to vary the high density behavior of EOS without affecting the bulk nuclear matter properties at saturation density. Mixed interaction terms involving $\rho$ mesons allow ones to significantly vary the density dependence of the symmetry energy coefficient which plays a crucial role in determining the cooling mechanism of a neutron star. We used the RMF model with three newly generated parameter sets DOPS1, DOPS2, and DOPS3  to calculate various EOSs composed of nucleons  and nucleons with quarks. The generated parameter sets of the model are calibrated by using the available  experimental data   \citep{Wang2017} on the total binding energy and the charge rms radii for a  few closed shell nuclei. We also used the value of neutron skin thickness for the $^{208}Pb$ nucleus in our calibration procedure. We employ our EOSs to study the structural properties of non-rotating  compact stars (CSs).
\par The manuscript  has been organized as, in section \ref{tf}, we described the theoretical
framework which is used to construct the various EOSs for pure nucleonic matter  and nucleonic with quark matter.
RMF model has been employed to describe the nucleonic phase  and the quark matter
phase has been obtained from the NJL model.
The coexisting phase of hybrid EOSs is obtained by using Glendenning 
construction based on Gibbs conditions of equilibrium.  In section \ref{nfmp}, we present our new parameterizations for RMF model. In section \ref{fnainm}, we present our results for finite nuclei and  bulk nuclear matter properties at saturation density. In this section, we also discuss the quality of fits to finite nuclei for the newly generated  parameterizations. In section \ref{eosansp}, we  present the set of EOSs generated and the  results for the various properties of non-rotating neutron stars are also discussed.
The summary is presented in section \ref{s}.

\section{Theoretical Formalism}  \label{tf}                                   
In this section, we discuss the theoretical  model employed to  calculate 
various EOSs of dense nuclear matter in different phases.
The newly generated parameter sets DOPS1, DOPS2,  and DOPS3 of the RMF model have been  successfully applied in describing the properties of finite nuclei and bulk nuclear matter 
at saturation density. These model parameters have been used 
to construct neutron stars and hybrid CSs.
The quark matter phase of the EOS has been calculated by using the NJL model.
The final hybrid EOS is comprised of two separate EOSs for 
each phase of matter, which are combined by utilizing a Glendenning phase transition
construction.   \citep{Glendenning1992,Glendenning2000}.
\subsection{Hadronic Equation of State}
 In the RMF model, the effective Lagrangian density consists of
self and mixed interaction terms for $\sigma$, $\omega$ and $\rho$ mesons up to
the quartic order in addition to the exchange interaction of baryons with $\sigma$, 
$\omega$ and  $\rho$ mesons. The $\sigma$, the $\omega$ , and the $\rho$ mesons are responsible for the ground state properties of the finite  nuclei ranging from low mass to heavy mass region in the periodic table. 
The mixed interactions terms containing the $\rho$-meson field enable us to vary
the density dependence of the symmetry energy coefficient and neutron skin thickness in
heavy nuclei over a wide range without affecting the other properties of the 
finite nuclei   \citep{Furnstahl2002,Sil2005}. In particular, the contribution from 
the self-interaction of $\omega$-meson  determines the
high density behavior of EOS and structure properties of CSs. 
  \citep{Dhiman2007,Muller1996}.  The inclusion of self-interaction of 
$\rho$-meson hardly affects the ground state properties of
heavy nuclei and compact stars   \citep{Muller1996}. 
The effective lagrangian density for the RMF model generally describes the interaction of the baryons via the exchange of $\sigma$, $\omega$ and $\rho$ mesons upto the quartic order. The lagrangian  density for  the RMF model   \citep{Dhiman2007,Raj2006} is given by
\begin{eqnarray}
\label{eq:lbm}
{\cal L} &=& \sum_{B} \overline{\Psi}_{B}[i\gamma^{\mu}\partial_{\mu}-
(M_{B}-g_{\sigma B} \sigma)-(g_{\omega B}\gamma^{\mu} \omega_{\mu}\nonumber\\&+&
\frac{1}{2}g_{\mathbf{\rho}B}\gamma^{\mu}\tau_{B}.\mathbf{\rho}_{\mu})]\Psi_{B}
+ \frac{1}{2}(\partial_{\mu}\sigma\partial^{\mu}\sigma-m_{\sigma}^2\sigma^2)\nonumber\\  &-&
\frac{\overline{\kappa}}{3!}
g_{\sigma N}^3\sigma^3-\frac{\overline{\lambda}}{4!}g_{\sigma N}^4\sigma^4  - \frac{1}{4}\omega_{\mu\nu}\omega^{\mu\nu}
+ \frac{1}{2}m_{\omega}^2\omega_{\mu}\omega^{\mu}\nonumber\\&+& \frac{1}{4!}\zeta g_{\omega N}^{4}(\omega_{\mu}\omega^{\mu})^{2}-\frac{1}{4}\mathbf{\rho}_{\mu\nu}\mathbf{\rho}^{\mu\nu}+\frac{1}{2}m_{\rho}^2\mathbf{\rho}_{\mu}\mathbf{\rho}^{\mu}\nonumber\\
&+&\frac{1}{4!}\xi g_{\rho N}^{4}(\mathbf{\rho}_{\mu}\mathbf{\rho}^{\mu})^{2} \nonumber\\
&+&  g_{\sigma N}g_{\omega N}^2\sigma\omega_{\mu}\omega^{\mu} \left(a_{1}+\frac{1}{2}a_{2}\sigma\right)\nonumber\\
&+&g_{\sigma N}g_{\rho
N}^{2}\sigma\rho_{\mu}\rho^{\mu} \left(b_{1}+\frac{1}{2}b_{2}
\sigma\right)\nonumber\\&+&\frac{1}{2}c_{1}g_{\omega N}^{2}g_{\rho N}^2\omega_{\mu}\omega^{\mu}\rho_{\mu}\rho^{\mu}
\end{eqnarray}
The energy density of the uniform matter  within the framework of RMF model is given by;
\begin{equation}
\label{eq:eden}
\begin{split}
{\cal E} & = \sum_{j=B,\ell}\frac{1}{\pi^{2}}\int_{0}^{k_j}k^2\sqrt{k^2+M_{j}^{*2}} dk\\
&+\sum_{B}g_{\omega B}\omega\rho_{B}
+\sum_{B}g_{\rho B}\tau_{3B}\rho_{B}\rho
+ \frac{1}{2}m_{\sigma}^2\sigma^2\\
&+\frac{\overline{\kappa}}{6}g_{\sigma N}^3\sigma^3
+\frac{\overline{\lambda}}{24}g_{\sigma N}^4\sigma^4
-\frac{\zeta}{24}g_{\omega N}^4\omega^4\\
&-\frac{\xi}{24}g_{\rho N}^4\rho^4
 - \frac{1}{2} m_{\omega}^2 \omega ^2
-\frac{1}{2} m_{\rho}^2 \rho ^2\\
&-a_{1} g_{\sigma N}
 g_{\omega N}^{2}\sigma \omega^2
 -\frac{1}{2}a_{2} g_{\sigma N}^2 g_{\omega N}^2\sigma^2 \omega^2\\
 &-b_{1}g_{\sigma N}g_{\rho N}^2 \sigma\rho^2
-\frac{1}{2} b_{2} g_{\sigma N}^2 g_{\rho N}^2\sigma^2\rho^2\\
 &- \frac{1}{2} c_{1} g_{\omega N}^2 g_{\rho N}^2
\omega^2\rho^2.\\
\end{split}
\end{equation}
The pressure of the uniform matter  is given by
\begin{equation}
\label{eq:pden}
\begin{split}
P & = \sum_{j=B,\ell}\frac{1}{3\pi^{2}}\int_{0}^{k_j}
\frac{k^{4}dk}{\sqrt{k^2+M_{j}^{*2}}} 
- \frac{1}{2}m_{\sigma}^2\sigma^2\\
&-\frac{\overline{\kappa}}{6}g_{\sigma N}^3\sigma^3 
-\frac{\overline{\lambda}}{24}g_{\sigma N}^4\sigma^4
+\frac{\zeta}{24}g_{\omega N}^4\omega^4\\
&+\frac{\xi}{24}g_{\rho N}^4\rho^4
  + \frac{1}{2} m_{\omega}^2 \omega ^2
+\frac{1}{2} m_{\rho}^2 \rho ^2 \\
& +a_{1} g_{\sigma N}
g_{\omega N}^{2}\sigma \omega^2
+\frac{1}{2} a_{2} g_{\sigma N}^2 g_{\omega N}^2\sigma^2 \omega^2\\
&+b_{1}g_{\sigma N}g_{\rho N}^2 \sigma\rho^2
+\frac{1}{2} b_{2} g_{\sigma N}^2 g_{\rho N}^2\sigma^2\rho^2\\
&+ \frac{1}{2} c_{1} g_{\omega N}^2 g_{\rho N}^2
\omega^2\rho^2.\\
\end{split}
\end{equation}
Here, the sum is taken over nucleons and leptons.
\par The composition of nuclear matter species i=n, p, e$^{-}$ and  $\mu^{-}$ at fixed baryon 
number density $\rho_{B}$=$\sum_{i}B_{i}\rho_{i}$ is determined in such
a way that the charge neutrality condition,
\begin{equation}
 \sum_{i}q_{i}\rho_{i}=0,
\end{equation}
and the chemical equilibrium conditions
\begin{equation}
\label{eq:mu}
 \mu_{i}=B_{i}\mu_{n}-q_{i}\mu_{e},
\end{equation}
are satisfied, where B$_{i}$ and q$_{i}$ denote baryon number
and electric charge of the species i.\\ 
\subsection{Quark Matter Equation of State}
\label{sec:model1}
We use the NJL model   \citep{Buballa2005,He2016}to calculate the EOS for the quark phase. By introducing the scalar-isovector and vector-isovector couplings, the largrangian of the three flavour NJL model can be written as 
\begin{eqnarray}
\label{sec:njl1}
 {\cal L}_{NJL} &=& \overline{q}(\iota \slashed{\partial} - \hat{m})q + \frac{G_{S}}{2}\sum_{a=0}^{8}[(\overline{q}\lambda_{a}q)^{2} + (\overline{q}\iota\gamma_{5}\lambda_{a}q)^{2}] \nonumber\\
 &+&\frac{G_{V}}{2}\sum_{a=0}^{8}[(\overline{q}\gamma_{\mu}\lambda_{a}q)^{2} + (\overline{q}\gamma_{5}\gamma_{\mu}\lambda_{a}q)^{2}]  \nonumber\\
 &-& K{det[\overline{q}(1+\gamma_{5})q] + det[\overline{q}(1-\gamma_{5})q]} \nonumber\\
 &+&{G_{IS}}\sum_{a=1}^{3}[(\overline{q}\lambda_{a}q)^{2} + (\overline{q}\iota\gamma_{5}\lambda_{a}q)^{2}]  \nonumber\\
 &+& {G_{IV}}\sum_{a=1}^{3}[(\overline{q}\gamma_{\mu}\lambda_{a}q)^{2} + (\overline{q}\gamma_{5}\gamma_{\mu}\lambda_{a}q)^{2}]
\end{eqnarray}

 Here q denotes the quark field with three flavours u,d and s, and three colours; $\hat{m}$=diag($m_{u},m_{d},m_{s}$) is the current quark mass matrix in three flavour space; $\lambda_{a}$ are the flavour SU(3) Gell-Mann matrices with $\lambda_{0}$ = $\sqrt{\frac{2}{3}}$I; $G_{S}$ and $G_{V}$ are the strength of the scalar and vector coupling, respectively; and K term represents the six-point Kobayashi-Maskawa-t'Hooft (KMT) interaction that breaks the axial $U(1)_{A}$  symmetry. Since the Gell-Mann matrics with a = 1-3 are identical to the Pauli matrics in u and d space, the last two terms represent the scalar-isovector and vector-isovector coupling breaking the SU(3) asymmetry while keeping the isospin symmetry, with $G_{IS}$ and $G_{IV}$  the corresponding coupling strength. In the present study, we employ the parameters $m_{u}$= $m_{d}$ = 3.6 MeV, $m_{s}$ = 87 MeV,
 $G_{S}{\Lambda}^{2}$ = 3.6, $K\Lambda^{5}$ = 8.9, and the cut off value in the momentum integral $\Lambda$ = 750 MeV which is taken from the references   \citep{Bratovic2013,Lutz1992,Buballa2005}. In the present work, we have used  vector coupling  $G_{V}$ = 0 in order to describe the astrophysical  constraints (Mass/Radius) of MSP 0740+6620, PSR J1614-2230 \cite{Cromartie2020,Demorest2010,Annala2018,Riley2021} as hybrid stars. However, the larger value of vector coupling $G_{V}$ can stiffen the resulting EOSs and may lead to different neutron star properties \cite{Peng2016}.
 In the NJL model, the quark masses are dynamically generated as solutions of the gap equation, obtained by imposing that the potential be stationary with respect to variations in the quark condensate <$\overline{q_{i}}q_{i}$>, thus finding 
 \begin{equation}
 \label{eq:gap}
  M_{i} = m_{i} - 2 G_{S}\sigma_{i} + 2 K\sigma_{j}\sigma_{k} - 2 G_{IS}\tau_{3i}(\sigma_{u}-\sigma_{d})
 \end{equation}
 where $\sigma_{i}$ = <$\overline{q_{i}}q_{i}$> stands for the quark condensate with (i,j,k) being any permutation number of (u,d,s), and $\tau_{3i}$ is the isospin quantum number of quark, i.e.,  $\tau_{3u}$ = 1,  $\tau_{3d}$ = -1 and  $\tau_{3s}$ = 0. As shown in the Eq. (\ref{eq:gap}), $\sigma_{d}$ and $\sigma_{s}$ contribute to the u quark mass through the KMT interaction as well as the scalar-isovector coupling, called the flavor mixing   \citep{Frank2003,Zhang2014} in the constituent quark mass.
 The quark condensate <$\overline{q_{i}}q_{i}$> and the quark number density $\rho_{i}$ are given respectively as below
 \begin{eqnarray}
 \label{eq:condensate}
  <\overline{q_{i}}q_{i}> = -2 N_{c}\int{\frac{d^{3}p}{(2\pi)^{3}}}\frac{M_{i}}{E_{i}} \\ \nonumber
  \rho_{i}= 2N_{c}\int_{0}^{\Lambda}\frac{d^{3}p}{(2\pi)^{3}}
 \end{eqnarray}
The above Eq.(\ref{eq:condensate}) has to be evaluated self-consistently with Eq.(\ref{eq:gap}), forming a set of six coupled equations for the constituent masses $M_{i}$. Once the self-consistent solutions are found, we can calculate the energy density and the pressure in the following form   \citep{He2016},
\begin{eqnarray}
\label{eq:energy}
 \epsilon_{NJL} &=& -2N_{c}\int_{0}^{\Lambda}\frac{d^{3}p}{(2\pi)^{3}}E_{i} + G_{S}({\sigma_{u}}^{2}+{\sigma_{d}}^{2}+{\sigma_{s}}^{2})\nonumber\\
 &-& 4 K \sigma_{u}\sigma_{d}\sigma_{s}+ G_{V}({\rho_{u}}^{2}+{\rho_{d}}^{2}+{\rho_{s}}^{2})\nonumber\\
 &+& G_{IS} (\sigma_{u}-\sigma{d})^{2} 
 + G_{IV} (\sigma_{u}-\sigma{d})^{2} - \epsilon_{0}
\end{eqnarray}
 In Eq. (\ref{eq:energy}),  $\epsilon_{0}$ is introduced to ensure that $\epsilon_{NJL}$ = 0 in the vacuum.
 The pressure for the cold quark matter can be calculated from the following equation
 \begin{equation}
  P = {\sum_{i=u,d,s}\mu_{i}\rho_{i}} - \epsilon_{NJL}
 \end{equation}
\subsection{Coexisting Phase}
We construct the EOS of coexisting phase (CP) made up of the hadron phase (HP) and quark matter phase for the hybrid compact star by implementing the Glendenning construction   \citep{Glendenning1992,
Glendenning2000}.
The evolution of coexisting phase is favored when the surface tension between  Coulomb interaction 
hadronic and quark matter is smaller   \citep{Alford2001} and negligible.
 The calculation of surface tension is very model dependent 
  \citep{Alford2001,Yasutake2014}. 
For the higher values of surface tension, the phase transition is sharp and this is to be
constructed with Maxwell construction and in the same way, low values of the 
phase transition is continuous and that is to  be constructed with Glendenning construction.
Since the value of surface tension is not
established yet, both of the methods of coexisting phase construction are equally
valid. But, we adopted the Glendenning construction based on the Gibbs condition of equilibrium.
The equilibrium chemical potential of
the coexisting phase corresponding to the intersection of the two surfaces representing
hadron and quark matter phase can be calculated  for mechanical 
and chemical equilibrium at zero temperature with the following relation, 
\begin{equation}
\label{eq:mp1}
P_{HP}(\mu_e,\mu_n)=P_{NJL}(\mu,\mu_e)=P_{CP},
\end{equation}
where $P_{HP}$, $P_{NJL}$, and $P_{CP}$ are the pressures of the hadron phase, quark  phase, and coexisting phase, respectively. In coexisting phase, we have considered chemical equilibrium
at the hadron-quark interface as well as inside each of the phases   \citep{Maruyama2007}, 
so that Eq.(\ref{eq:mu}) implies
\begin{equation}
 \mu_{u}+\mu_{e}=\mu_{d}=\mu_{s},
\end{equation}
\begin{equation}
 \mu_{p}+\mu_{e}=\mu_{n}=\mu_{u}+2\mu_{d}.
\end{equation}
In the coexisting phase, the local charge neutrality condition is replaced by the 
global charge neutrality which means that both hadron and quark matter is 
allowed to be charged separately.
The condition of the global charge neutrality determines the volume fraction $\chi$ of the quark  phase and that  can be obtained by using, 
\begin{equation}
\chi\rho_{c}^{NJL}+(1-\chi)\rho_{c}^{HP}=0,
\end{equation}
where, the $\rho_{c}^{NJL}$ and $\rho_{c}^{HP}$ are the charge densities of the NJL phase and hadron phase of dense matter, respectively. The value of the $\chi$ increases
from zero in the pure hadron phase to $\chi=1$ in the pure quark phase.
The energy density ${\cal E}_{CP}$ and the baryon density $\rho_{CP}$ of the 
coexisting phase can be calculated as,
\begin{equation}
\label{eq:mp2}
{\cal E}_{CP}=\chi{\cal E}_{NJL}+(1-\chi){\cal E}_{HP},
\end{equation}
\begin{equation}
\rho_{CP}=\chi\rho_{NJL}+(1-\chi)\rho_{HP}.
\end{equation}
The coexisting phase of EOSs has been computed by employing the procedure explained above and Eqs.(\ref{eq:mp1}-\ref{eq:mp2}).
\subsection{Tidal deformability}
The tidal influences of its companion in BNS system will deform  CS in binary system and, the resulting change in the gravitational potential modifies 
the BNS orbital motion and its corresponding gravitational wave (GW) signal. This effect on GW phasing can be parameterized by the dimensionless tidal deformability parameter,
$\Lambda_i = \lambda_i/M_i^5,$ i = 1, 2.  For each CS, its quadrupole moment ${\cal{Q}}_{j,k}$ must be related to the tidal field ${\cal{E}}_{j,k}$ caused by its companion
as, ${\cal{Q}}_{j,k} = -\lambda {\cal {E}}_{j,k}$, where, $j$ and $k$ are spatial tensor indices.
The dimensionless tidal deformability
parameter  $\Lambda$  of a static, spherically symmetric compact star
depends on the neutron star compactness parameter C and a dimensionless quadrupole Love number k$_{2}$ as, $\Lambda$=(2k$_{2}/{3)C^{-5}}$. The $\Lambda$ critically parameterizes 
the deformation of CS under the given tidal field, therefore it should depend on the EOS of nuclear dense matter. When the orbital separation is  very small at the 
frequencies in the BNS systems, the tidal corrections are added to the tidal energy and luminosity linearly to the point-particle energy and luminosity.  The leading-order tidal 
corrections are Newtonian effects and, are known as 5PN (Post Newtonian) and next-to-leading-order 6PN corrections to the energy and luminosity   \citep{Favata2014,Wade2014}.
These leading-order tidal corrections are  required to be included in the waveform model employed for analysis  of GW signals from advanced LIGO and Virgo GW detectors at the high frequencies, as discussed for the various waveforms by Abbott et al,  \citep{Abbott2019}. 

To measure the Love number k$_{2}$ along with the evaluation of the TOV 
equations we have to compute y$_{2}$ = y(R) with 
initial boundary condition y(0) = 2 from the first-order
differential equation   \citep{Hinderer2008,Hinderer2009,Hinderer2010,Damour2010} simultaneously, 
 \begin{eqnarray}
 \label{y}
   y^{\prime}=\frac{1}{r}[-r^2Q-ye^{\lambda}\{1+4\pi Gr^2(P-{\cal{E}})\}-y^{2}], 
 \end{eqnarray}
where  Q $\equiv$ 4$\pi$Ge$^{\lambda}$(5${\cal{E}}$+9P+$\frac{{\cal{E}}+P}{c_{s}^2})$
 -6$\frac{e^{\lambda}}{r^2}$-$\nu^{\prime^2}$ and       
  e$^{\lambda} \equiv(1-\frac{2 G m}{r})^{-1}$ and, $\nu^{\prime}\equiv$ 2G e$^{\lambda}$ 
       ($\frac{m+4 \pi P r^3}{r^2}$).
First, we get the solutions of Eq.(\ref{y}) with boundary condition, y$_{2}$ = y(R),
then the electric tidal Love
number k$_{2}$ is calculated from the expression as,
\begin{eqnarray}
 k_{2}=\frac{8}{5}C^{5}(1-2C)^{2}[2C(y_{2}-1)-y_{2}+2]\{2C(4(y_{2}+1)C^4\nonumber\\
 +(6y_{2}-4)C^{3}
 +(26-22y_{2})C^2+3(5y_{2}-8)C-3y_{2}+6)\nonumber\\
 -3(1-2C)^2(2C(y{_2}-1)-y_{2}+2) \log(\frac{1}{1-2C})\}^{-1}.\nonumber\\
\end{eqnarray}

\section{New RMF model parameterization} \label{nfmp}
There are several relativistic mean field models in which energy density functional consists of
 nonlinear $\sigma$, $\omega$ and $\rho$ terms and mixed interaction terms.  These models  are used to construct the EOSs composed of nucleonic matter    \citep{Dutra2014} and nucleonic along with hyperonic matter   \citep{Dutra2016,Fortin2016} and accosted with the constraints of nuclear matter properties and astrophysical observations of CS masses   \citep{Antoniadis2013,Riley2019,Raaijmakers2019}. Only RMF models BSR   \citep{Dhiman2007} with $\zeta = 0$  and NL3$\omega\delta$   \citep{Horowitz2001} can sustain the condition of maximum mass M $\geq$ 2.0M$_\odot$ when hyperons are included in the EOSs with appropriate meson-hyperon couplings, otherwise, the inclusion of hyperons may lead for the famous hyperon puzzle. However, many RMF models   \citep{Lourenco2019} without the inclusion of hyperons satisfy the constraints of astrophysical observations obtained from binary neutron star merger event GW170817. In the present work, we search for the best fit parameters of RMF model by using simulated anealing method to minimise the  $\chi^{2}$   \citep{Burvenich2004,Kirkpatrick1984} which is given by 
 \begin{equation}\label{chi2}
 \chi^2 =  \frac{1}{N_d - N_p}\sum_{i=1}^{N_d}
\left (\frac{ M_i^{exp} - M_i^{th}}{\delta_i}\right )^2 
\end{equation}
where, $N_d$ is the number of  experimental data points and $N_p$ the
number of fitted  parameters. The $\delta_i$ stands for theoretical error \cite{Dobaczewski2014}
and $M_i^{exp}$ and $M_i^{th}$ are the experimental and the corresponding
theoretical values, respectively, for a given observable. Since $M_i^{th}$ in Eq. (\ref{chi2}) are calculated by using RMF model, the value of $\chi^2$ depends on the parameters appearing in Eq. (\ref{eq:lbm}). The theoretical errors $\delta_i$ in Eq. (\ref{chi2}) are taken to be 1.0 MeV for total binding energies, 0.02 fm for the charge rms radii and 0.005 fm for the neutron skin thickness.
Three new parameter sets namely DOPS1, DOPS2, and DOPS3 have been generated by including  all possible self and mixed interaction terms for $\sigma$, $\omega$ and $\rho$ mesons up to quartic order  for a fixed value of  $\omega$ meson self-coupling parameter $\zeta$ = 0.00, 0.01 and 0.02.  The remaining coupling parameters are determined by fitting the RMF results to the available experimental data for  total binding energies for  $^{16,24}$O, $^{40,48}$Ca, $^{56,78}$Ni,
$^{88}$Sr, $^{90}$Zr, $^{100,116,132}$Sn, and $^{208}$Pb nuclei and  charge
rms radii for $^{16}$O, $^{40,48}$Ca, $^{56}$Ni, $^{88}$Sr, $^{90}$Zr,
$^{116}$Sn, and $^{208}$Pb nuclei as per the available experimental data    \citep{Wang2017}. In addition, we also fit the value of
neutron skin thickness for the  $^{208}$Pb nucleus which is a very important physical observable. Recently extracted values of
neutron skin thickness for the  $^{208}$Pb nucleus from  isospin diffusion
data lie within $0.16-0.33{ fm}$ indicating large uncertainties
  \cite{Chen2005,Fattoyev2020,Abrahamyan2012}. It is also shown in ref. \cite{Piekarewicz2004} that neutron skin thickness of $\approx$ 0.18 fm in the  $^{208}Pb$ nucleus is required to adequately reproduce the centroid energies of isoscalar giant monopole and isovector giant dipole resonances. We  include in our fit, the value of neutron skin thickness $\Delta$r=0.18 fm for  the $^{208}Pb$ nucleus to constrain the linear density dependence of the symmetry energy coefficient.
 The DOPSs parameter sets have been generated for a fixed value of  $\omega$-meson self-coupling parameter $\zeta$ = 0.00, 0.01, and 0.02 in the light of the recent observation of GW190814, GW170817, and PSR J0740+6620. The coupling parameter  $\zeta$ affects the high density behavior of the EOS. A large value of $\zeta$ makes the EOS softer and a smaller value stiffens the EOS.  The value of maximum mass for GW190814 event lies in the range 2.50-2.67 $M_{\odot}$ \cite{Abbott2020}, which requires stiff EOS and hence a very small value of $\zeta$ which we have taken equal to zero for DOPS1 parameterization. The astrophysical events  GW170817 and PSR J0740+6620 have maximum mass $\approx$ 2 $M_{\odot}$ and require relatively softer EOSs. For this,  we have fixed the value of $\zeta$ equal to 0.01 and 0.02 for DOPS2 and DOPS3 parameter sets respectively.
 The $\rho$ meson self interaction has not been included as it hardly affects the properties of finite nuclei and neutron star   \citep{Muller1996}.\\
 \begin{table*}
\centering
\caption{\label{tab:table1}
Newly generated parameter sets DOPS1, DOPS2 and DOPS3 for  the Lagrangian of RMF model as given
in Eq.(\ref{eq:lbm}).
The parameters $\overline{\kappa}$, $a_1$, and $b_1$ are
 in  fm$^{-1}$. The masses  $m_{\sigma}$, $m_{\omega}$ and $m_{\rho}$  are in  MeV. The mass for nucleon  is taken as $M_N=939 MeV$.
The values of $\overline{\kappa}$, $\overline{\lambda}$, $a_1$, 
$a_2$, $b_1$, 
$b_2$, 
and $c_1$ are multiplied by $10^{2}$. The parameter sets NL3, FSUGarnet, IOPB-1 and Big Apple are also presented.}
\vskip 1cm
\begin{tabular}{cccccccc}
\hline
\hline
\multicolumn{1}{c}{${\bf {Parameters}}$}&
\multicolumn{1}{c}{{\bf DOPS1}}&
\multicolumn{1}{c}{{\bf DOPS2}}&
\multicolumn{1}{c}{{\bf DOPS3}}&
\multicolumn{1}{c}{{\bf NL3}}&
\multicolumn{1}{c}{{\bf FSUGarnet}}&
\multicolumn{1}{c}{{\bf IOPB-1}}&\multicolumn{1}{c}{{\bf Big Apple}}\\
\hline
${\bf g_{\sigma}}$ &10.20651&10.67981& 10.51853&10.21743 &10.50315&10.41851&9.67810\\
${\bf g_{\omega}}$ &12.87969&14.12312 &13.53456 & 12.86762&13.69695&13.38412&12.33541\\
${\bf g_{\rho}}$ &14.13399 &14.89809 &14.64808 &8.94880 &13.87880 &11.11560&14.14256\\
${\bf \overline {\kappa}}$&2.62033&3.00823 &1.97233 &1.95734 &1.65229 &1.85581&2.61776\\
${\bf \overline {\lambda}}$&-1.67616&-0.07894 &-0.11438 &-1.59137&-0.35330& -0.75516&-2.16586\\
${\bf{\zeta}}$&0.00000&0.01000 &0.02000 &0.00000&0.23486& 0.017442&0.000699\\
${\bf a_1}$& 0.02169&0.08832 &0.01911 &0.00000 &0.00000 &0.00000&0.00000\\
${\bf  a_2}$ &0.01785&0.04637 &0.02699 &0.00000 &0.00000 &0.00000&0.00000\\
${\bf b_1}$ &0.73554 &0.79273 &0.77259 &0.00000 &0.00000 &0.00000&0.00000\\
${\bf b_2}$ &0.98545&0.98986 &0.89590 &0.00000 &0.00000 &0.00000&0.00000\\
${\bf c_1}$ &0.86916&0.63710&0.70353&0.00000 &8.60000 &4.80000&9.40000\\
${\bf m_{\sigma}}$&503.61992 &492.84789 &502.37396 &508.19400 &496.93900 &500.48700&492.73000\\
${\bf m_{\omega}}$&782.50000 &782.50000 &782.50000 &782.50100 &782.50000 &782.18700&782.50000\\
${\bf m_{\rho}}$&770.00000&770.00000 &770.00000 &763.00000 &763.00000 &762.46800&763.00000\\
\hline
\hline
\end{tabular}
\end{table*}

In Table  \ref{tab:table1}, the newly generated parameter sets DOPS1, DOPS2 and DOPS3 are listed. We also display  the value of parameters for  NL3   \citep{Lalazissis1997},  FSUGarnet   \citep{Chen2015}, IOPB-1   \citep{Kumar2018} and Big Apple   \citep{Fattoyev2020}.
The effective field theory imposes the condition of naturalness   \citep{Furnstahl1997} on the parameters or expansion coefficients appearing in the energy density functional Eq. (\ref{eq:eden}). According to naturalness, the coefficients of various terms in energy density functional should be  of same size when expressed in appropriate dimensionless ratio. The dimensionless ratios are obtained by dividing Eq. (\ref{eq:eden}) by $M^{4}$ and expressing each term in powers of $\frac{g_{\sigma}\sigma}{M}$, $\frac{g_{\omega}\omega}{M}$ and 2$\frac{g_{\rho}\rho}{M}$. This means that the dimensionless ratios ${\frac{1}{2C_{\sigma}^{2}M^{2}}}$,${\frac{1}{2C_{\omega}^{2}M^{2}}}$, ${\frac{1}{8C_{\rho}^{2}M^{2}}}$, ${\frac{\overline{\kappa}}{6M}}$, ${\frac{\overline{\lambda}}{24M}}$, ${\frac{{\zeta}}{24}}$, ${\frac{a_1}{M}}$, ${\frac{a_2}{2}}$, ${\frac{b_1}{4M}}$, ${\frac{b_2}{8}}$ and ${\frac{c_1}{8}}$ should be roughly of same size, where ${c_{i}}^{2}=\frac{{g_{i}}^{2}}{{m_{i}}^{2}}$, i denotes $\sigma$, $\omega$ and $\rho$ mesons.
\begin{table*}
\centering
\caption{\label{tab:table2}
The values of parameters expressed as dimensionless ratios corresponding to naturalness behavior. All  values have been  multiplied by $10^{3}$.} 
\vskip 1cm
\begin{tabular}{cccccccc}
\hline
\hline
\multicolumn{1}{c}{${\bf {Parameters}}$}&
\multicolumn{1}{c}{{\bf DOPS1}}&
\multicolumn{1}{c}{{\bf DOPS2}}&
\multicolumn{1}{c}{{\bf DOPS3}}&
\multicolumn{1}{c}{{\bf NL3}}&
\multicolumn{1}{c}{{\bf FSUGarnet}}&
\multicolumn{1}{c}{{\bf IOPB-1}}&\multicolumn{1}{c}{{\bf Big Apple}}\\
\hline
${\bf{\frac{1}{2C_{\sigma}^{2}M^{2}}}}$ &1.3806&1.2076& 1.2935&1.4028 &1.2690&1.3086&1.4698\\
${\bf{\frac{1}{2C_{\omega}^{2}M^{2}}}}$ &2.0931&1.7407 &1.8959& 2.0970&1.8508&1.9383&2.2819\\
${\bf{\frac{1}{8C_{\rho}^{2}M^{2}}}}$ &0.4207&0.3787&0.3917 &1.0306&0.4278&0.6670 &0.4121\\
${\bf{\frac{\overline{\kappa}}{6M}}}$&0.9177&1.0536&0.6908 &0.6855 &0.5787 &0.6499&0.9168\\
${\bf{\frac{\overline{\lambda}}{24M}}}$&-0.6984&-0.0329 &-0.0476&-0.6630&-0.1472& -0.3146&-0.9024\\
${\bf{\frac{{\zeta}}{24}}}$&-&0.4166 &0.8333 &-&0.9785& 0.7267&0.0291\\
${\bf{\frac{a_1}{M}}}$ &0.2169&0.8832 &0.1911 &- &- &-&-\\
${\bf{\frac{a_2}{2}}}$& 0.0893&0.2318&0.1349 &- &- &-&-\\
${\bf{\frac{b_1}{4M}}}$ &1.8388&1.9818 &1.9315 &- &- &-&-\\
${\bf{\frac{b_2}{8}}}$ &1.2318 &1.2373&1.1198 &- &- &-&-\\
${\bf{\frac{c_1}{8}}}$  &1.0864&0.7964&0.8794 &- &10.7500 &6.0000&11.7500\\
\hline
\hline
\end{tabular}
\end{table*}
In Table  \ref{tab:table2}, we present the overall naturalness behavior of various parameterizations i.e. the  value of these parameters when expressed in dimensionless ratios as shown just above. We also display the corresponding values for NL3, FSUGarnet, IOPB-1, and Big Apple parameter sets. It is clear from the table that the DOPS1, DOPS2 , and DOPS3 parameterizations closely favor the naturalness behavior. It can also be seen from   table \ref{tab:table2}, that the value of parameter $c_{1}$ (mixed interaction term of $\omega^{2}\rho^{2}$) is very large and equal to  10.75, 6.0 and 11.75 for FSUGarnet, IOPB-1 and Big Apple parameterizations respectively when expressed in appropriate dimensionless ratio. The large value of $c_{1}$ gives rise to the deviation from the naturalness behavior and this deviation might be attributed to the fact of not including all possible mixed interaction terms of $\sigma$, $\omega$ and $\rho$ mesons in these respective parameterizations,unlike DOPSs parameterizations. As far as NL3 parameterization is concerned, the naturalness behavior is favored very well but it does not include any cross interaction terms of sigma, omega, and rho mesons which are very important for constraining the symmetry energy and its density dependence.  DOPS1, DOPS2 , and DOPS3 parameterizations show   better naturalness behavior as compared to other parameterizations displayed in the table. The naturalness behavior of parameters can be further improved by considering the next higher order terms containing the gradient of fields   \citep{Furnstahl1997}.
\section{Finite nuclei and Infinite Nuclear matter} \label{fnainm}  In this section, we discuss our results for finite nuclei and infinite nuclear matter.
The newly generated parameterizations DOPS1, DOPS2 and DOPS3 give equally good fit to the properties of finite nuclei.
\begin{figure}
\includegraphics[trim=0 0 0 0,clip,scale=0.38]{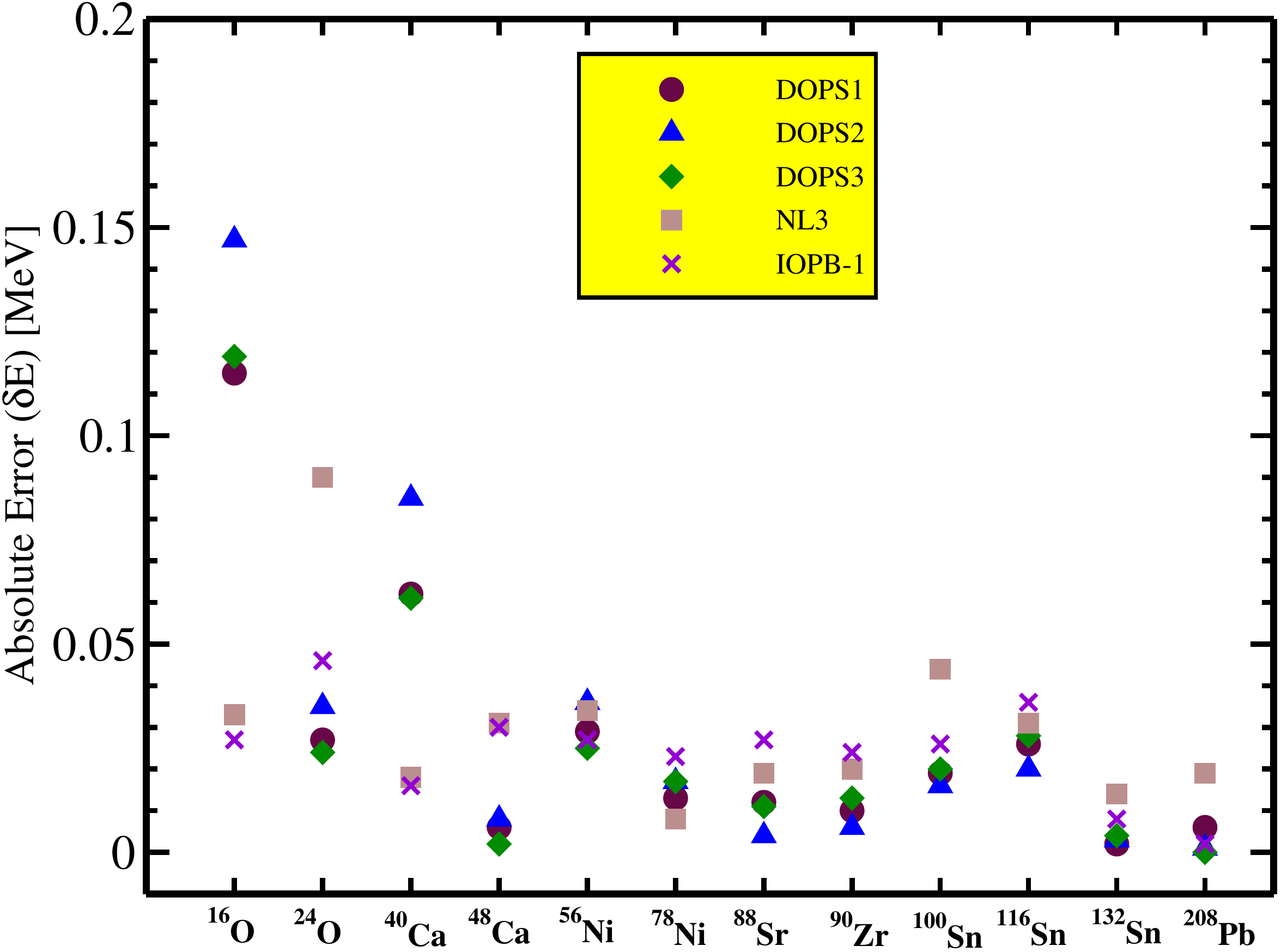}
\caption{\label{error_be} (Color online) Absolute error in the  binding energy per nucleon ($\delta E$) plotted against the mass number (A) for newly generated parameter sets DOPS1,DOPS2 and DOPS3. For comparison, the values of $\delta E$ obtained with parameters NL3 and IOPB-1 are also displayed. }
\end{figure}
In Fig. (\ref{error_be}), we display the value of absolute error  in   binding energy per nucleon which is defined as,
\begin{equation}
\delta E =  {|BE^{exp}-BE^{th}}|
\end{equation}
Here, $BE^{exp}$ and $BE^{th}$ are the experimental and theoretical values for the binding energy per nucleon respectively. Results for $\delta E$   are 
calculated for DOPSs parameterizations. The mean absolute errors in the binding energy per nucleon calculated  with the DOPS1, DOPS2, and DOPS3  parameterizations for the finite nuclei used in the fit are 0.027, 0.031 , and 0.027 MeV respectively. We also display  similar results for NL3, IOPB-1 parameter sets. It is evident that binding energies obtained using DOPSs parameterizations are in good agreement with the available experimental data   \citep{Wang2017}.
\begin{figure}
\includegraphics[trim=0 0 0 0,clip,scale=0.40]{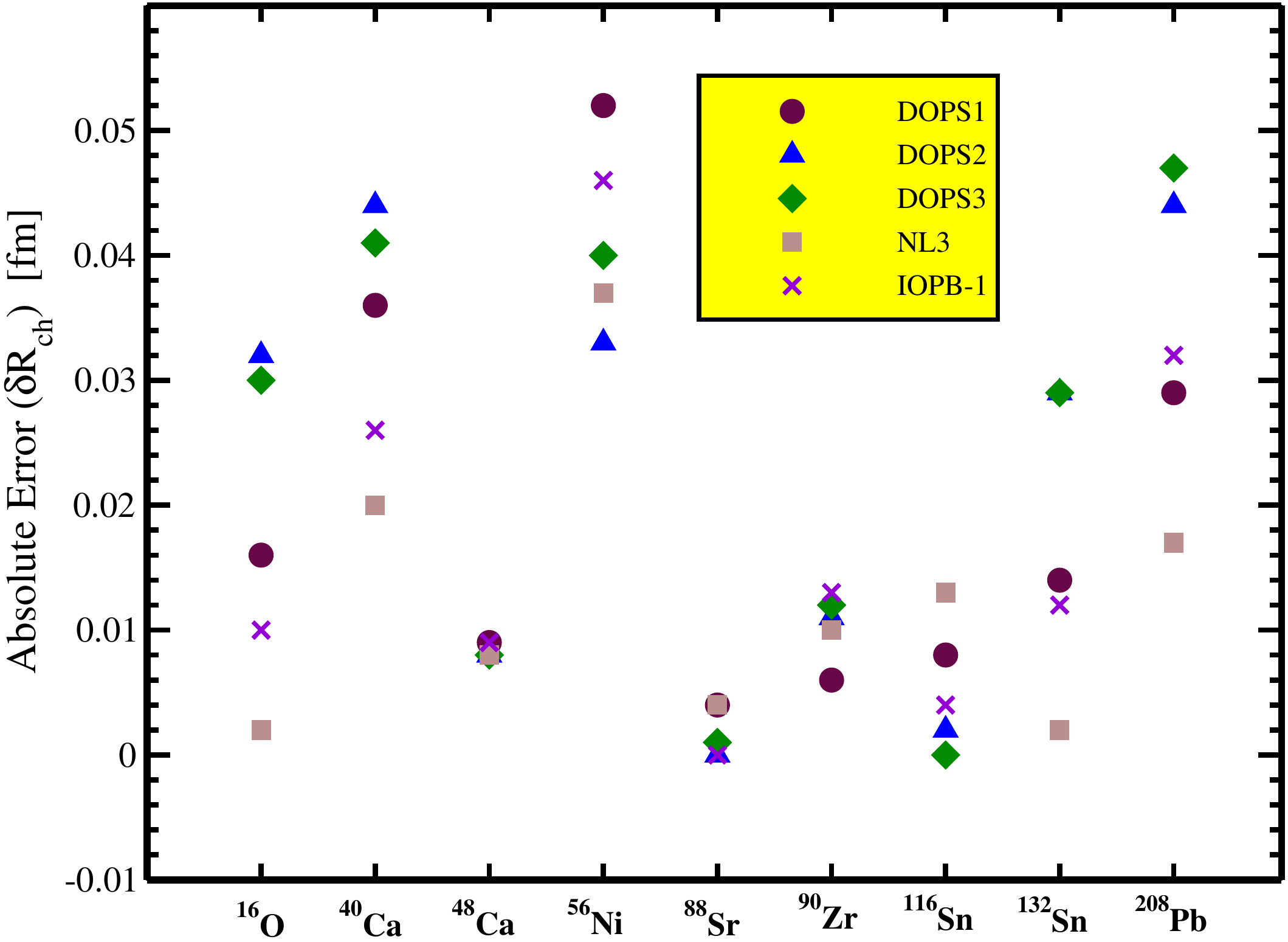}
\caption{\label{error_ch} (Color online) Absolute error in the charge root mean square radii ($\delta R_{ch}$) plotted against the mass number (A) for newly generated parameter sets DOPS1,DOPS2 and DOPS3. For comparison, the values  obtained with parameters NL3 and IOPB-1 are also displayed.}
\end{figure}
In Fig. (\ref{error_ch}), we present our results for absolute error $\delta R_{ch} =   {|R_{ch}^{exp}-R_{ch}^{th}}|$ for charge rms radii and also compare them with NL3 and IOPB-1 parameter sets. The value of charge rms radii calculated for various parameterizations displayed in Fig. (\ref{error_ch}) are more or less same. The mean absolute error in the charge rms radii for  DOPS1, DOPS2 and DOPS3  parameterizations for the finite nuclei used in the fit are 0.019, 0.022 and 0.023 fm respectively. \\
We have also calculated the rms errors in the total binding energy and charge radii for the nuclei considered in our fit.  The root mean square (rms) errors in total binding energy for all the nuclei considered in our fit are found to be  1.58, 1.63, 1.61, 2.41 , and 1.93  MeV for DOPS1, DOPS2, DOPS3, NL3 , and IOPB-1 parameterizations respectively. Similarly,  the root mean square (rms) errors in  charge radii for all nuclei taken in our fit are 0.020, 0.023, 0.024, 0.020, and 0.022 fm for DOPS1, DOPS2, DOPS3, NL3 , and IOPB-1 parameter sets respectively.
\begin{table*}
\centering
\caption{\label{tab:table3}
The SNM properties at saturation density for the parameter sets DOPS1, DOPS2 and DOPS3 are compared with that obtained using NL3, FSUGarnet, IOPB-1 and Big Apple parameter sets. $\rho_{0}$,  E/A,  K,  J,  L and $M^{*}/M$ denotes the saturation density, Binding Energy per nucleon, Nulcear Matter incompressibility coefficient, Symmetry Energy coefficient, density dependence  of symmetry energy and ratio of effective nucleon mass to the nucleon mass respectively.}  
\vskip 1cm
\begin{tabular}{cccccccc}
\hline
\hline
\multicolumn{1}{c}{${\bf {Parameters}}$}&
\multicolumn{1}{c}{{\bf DOPS1}}&
\multicolumn{1}{c}{{\bf DOPS2}}&
\multicolumn{1}{c}{{\bf DOPS3}}&
\multicolumn{1}{c}{{\bf NL3}}&
\multicolumn{1}{c}{{\bf FSUGarnet}}&
\multicolumn{1}{c}{{\bf IOPB-1}}&\multicolumn{1}{c}{{\bf Big Apple}}\\
\hline
${\bf{\rho_{0} ~(fm^{-3})}}$ &0.150&0.148 &0.148&0.148 &0.153&0.149&0.155\\
${\bf E/A ~(MeV)}$ &-16.073&-16.073&-16.037 & -16.248&-16.229&-16.099&-16.339\\
${\bf K ~(MeV)}$ &231.204&232.733&227.500&271.565 &229.623&222.571&227.093\\
${\bf J ~(MeV)}$&31.894&31.767 &31.843 &37.400&30.983&33.303&31.410\\
${\bf L ~(MeV)}$&65.590&66.018 &66.743&118.563&50.925& 63.850&40.339\\
${\bf ~M^{*}/M}$&0.604&0.611 &0.605 &0.595&0.578& 0.595&0.608\\
\hline
\hline
\end{tabular}
\end{table*}
In Table \ref{tab:table3}, we present our results for the symmetric nuclear matter (SNM) properties such as binding energy per nucleon (E/A), incompressibility (K) ,  symmetry energy coefficient (J), density dependence of symmetry energy (L) and the ratio of effective mass to the mass of nucleon ($M^{*}/M$) at the saturation density ($\rho_{0}$). These properties are very important for constructing the EOS for nuclear matter. The value of E/A is $\approx$ -16 MeV for all DOPSs parameterizations. For all newly generated parameterizations, the  value of J and L are consistent with the constraints from observational analysis J = 31.6 $\pm$ 2.66 MeV and L = 58.9 $\pm$ 16 MeV   \citep{Li2013}. The value of K lies in the range 227.5 - 232.733 MeV which is in agreement with the value of K = 240 $\pm$ 20 MeV determined from isoscalar giant monopole resonance (ISGMR) for  $^{90}Zr$ and $^{208}Pb$ nuclei       \citep{Colo2014,Piekarewicz2014}. The ratio of effective mass to the nucleon mass is found to be similar for all DOPSs parameterizations as shown in Table \ref{tab:table3}. The SNM properties calculated with NL3, FSUGarnet, IOPB-1 , and Big Apple are also shown for comparison.
\begin{figure}
\includegraphics[trim=0 0 0 0,clip,scale=0.55]{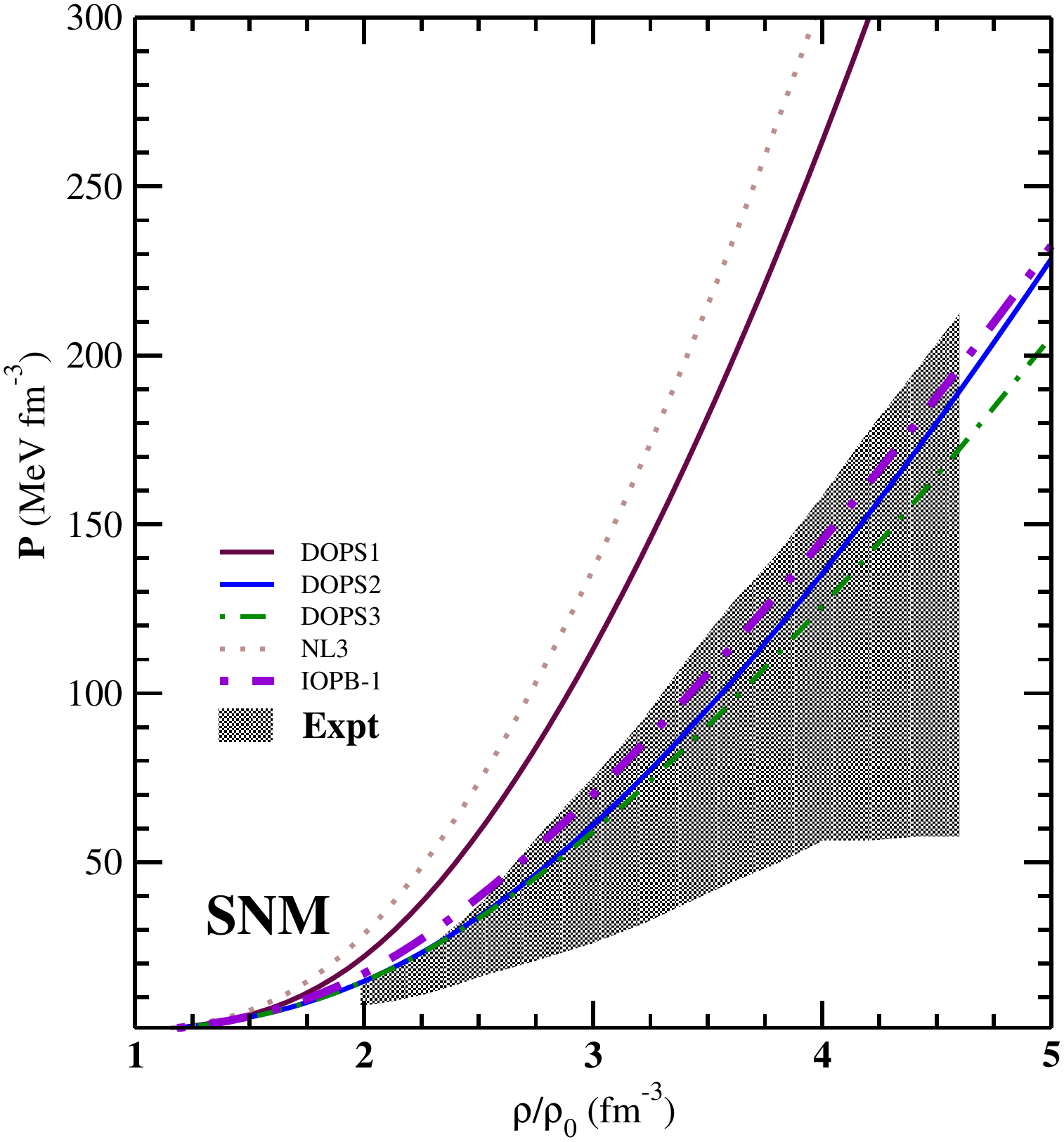}
\caption{\label{snm} (Color online) Variation of Pressure as a function of baryon density  for symmetric nuclear matter  (SNM) computed with DOPS1, DOPS2 and DOPS3 parameterizations along with NL3 and IOPB-1. The shaded region represents the experimental data taken from the reference   \citep{Danielewicz2002}.}
\end{figure}
\begin{figure}
\includegraphics[trim=0 0 0 0,clip,scale=0.52]{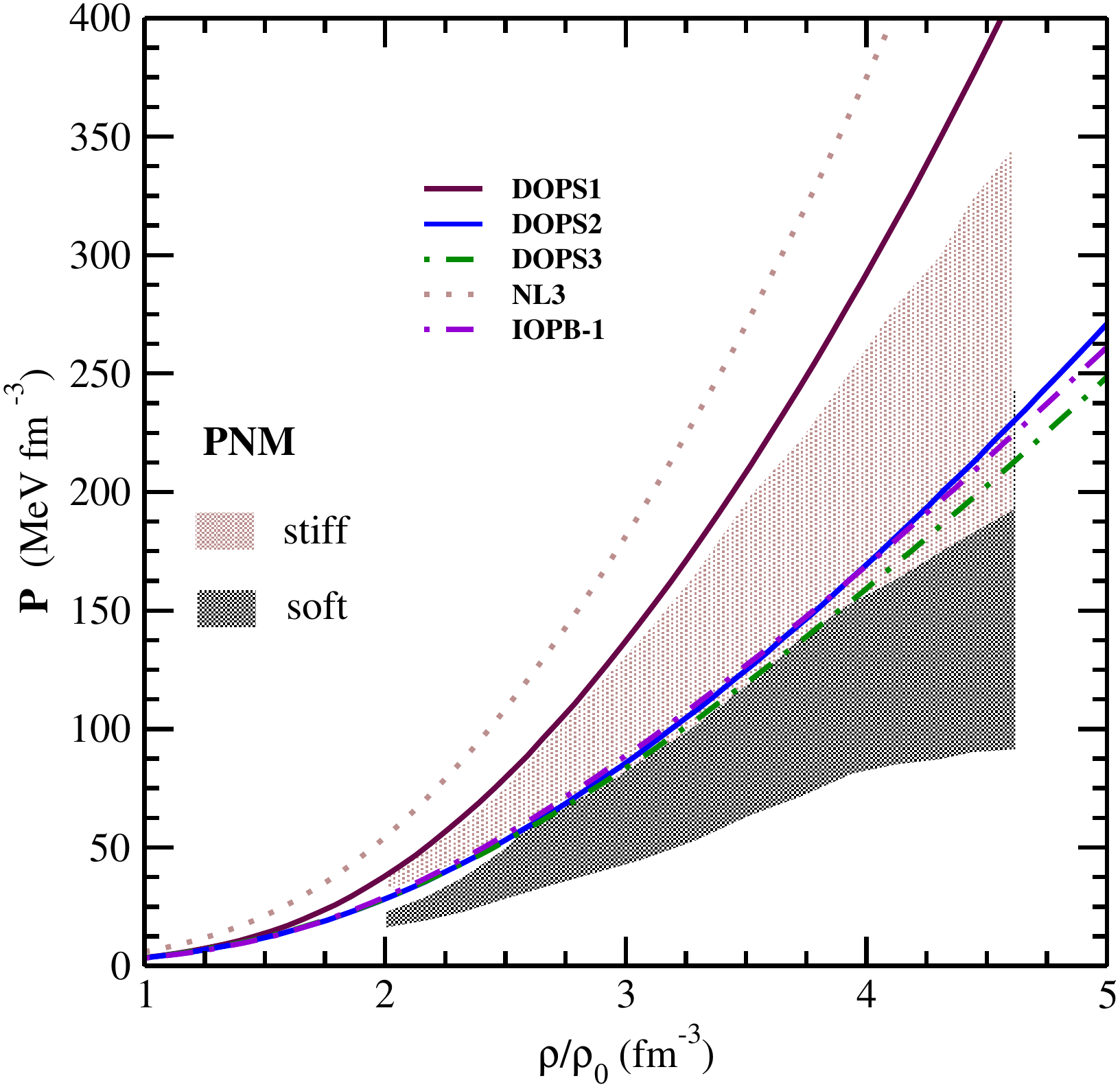}
\caption{\label{pnm} (Color online) Variation of Pressure as a function of baryon density  for pure neutron matter  (PNM) computed with DOPS1, DOPS2 and DOPS3 parameterizations along with NL3 and IOPB-1. The shaded region represents the experimental data taken from the reference   \citep{Danielewicz2002}.}
\end{figure}
In Fig. (\ref{snm} and \ref{pnm}), we plot the EOS i.e. pressure as a function of baryon density ($\frac{\rho}{\rho_{0}}$) for SNM and pure neutron matter (PNM) using DOPS1,DOPS2 and DOPS3 parameterizations which is in good agreement and lie in the allowed region  with the EOS extracted from the analysis of particle flow in heavy-ion collision   \citep{Danielewicz2002}.
 These results are also compared with the NL3 and IOPB-1 parameterizations. It can be easily seen that the EOSs for SNM and PNM obtained from DOPS1 and NL3 parameterizations are very stiff and are ruled out by heavy ion collision data. The stiffness of the EOSs for DOPS1 and NL3 parameter sets may be due to the fact that the  coupling parameter $\zeta$  which  varies the high density behavior of EOS is taken to be equal to  zero. The stiff EOS obtained  by DOPS1 is required to account for the predicted  supermassive neutron star in GW190814 event.
 The EOSs calculated using DOPS2 and DOPS3 parameter sets are much softer  and lie in the allowed region of  heavy ion collision data   \citep{Danielewicz2002}. The softness of EOSs is attributed to the large value of $\zeta$. 
 \\
 DOPS1 gives Stiffest EOS among DOPS parameter sets and hence a large value of energy density and pressure at a given baryon density. The parameter sets  DOPS2 and DOPS3 give relatively softer EOSs and comparatively smaller value of pressure and energy density. Due to this fact, the lines for model  DOPS1 are  so much different than that of DOPS2 and DOPS3 model parameterization as shown in Fig. (\ref{snm} and \ref{pnm}).
\section{Equation of State and Neutron Star Properties} \label{eosansp}
Here we discuss the results for the properties of non-rotating neutron stars for a set of EOSs obtained using different parameterizations in the hadronic phase, uds quark phase , and coexisting phase.  We employed the Baym-Pethick-Sutherland (BPS)   \citep{Baym1971} EOS for low density regime from outer crust baryon density ($\rho$  = 6.3 $\times 10^{-12}$) up to the pasta phase ($\rho$ = $9.4 \times 10^{-2}$). The crust region and the core region of the EOSs have been matched by using the cubic interpolation method that offers true continuity between the crust and the core.

\begin{table*}
 \caption{\label{tab:table4}The properties  of non-rotating compact stars for the various EOSs along with their particle composition computed with the newly generated parameter sets are presented. Results are also displayed for other parameter sets. M$_G(M_\odot)$ and  R$_{max}$  denote the Maximum Gravitational mass and  radius corresponding to the maximum mass  of the non-rotating compact stars respectively. The 
 values for R$_{1.4}$ and $\Lambda_{1.4}$ denote radius and  dimensionless tidal deformability at 1.4M$_\odot$.
}
 \begin{tabular}{|c|c|c|c|c|c|c|c|}
\hline
\bf{No.} &\bf{EOS}&\bf{Particle}& \bf{M(M$_{\odot}$)}& \bf{R$_{max}$ } & \bf{R$_{1.4}$}&\bf{$\Lambda_{1.4}$}\\
 & &\bf{composition} & & (km) &(km) & \\
 \hline
1 &DOPS1&n,p,e,$\mu$&2.57&12.36&13.61&627.37\\
2 &DOPS2&n,p,e,$\mu$&2.12&11.56&13.24&546\\
3 &DOPS3&n,p,e,$\mu$&2.05&11.61&13.25& 563.2\\
4 &DOPS1Q&n,p,e,$\mu$,u,d,s&2.23&13.22&13.60&637\\
5&DOPS2Q&n,p,e,$\mu$,u,d,s&1.95&12.39&13.25&527.3\\
6&DOPS3Q&n,p,e,$\mu$,u,d,s&1.91&12.35&13.26&547.2\\
7&NL3&n,p,e,$\mu$&2.73&13.10&14.65&1234.8\\
8&FSUGarnet&n,p,e,$\mu$&2.06&11.70&12.86&624.8\\
9&IOPB-1&n,p,e,$\mu$&2.16&12.22&14.09&833\\
10&Big Apple&n,p,e,$\mu$&2.6&12.41&12.96&717.3\\

\hline
\end{tabular}
\end{table*}
In Table \ref{tab:table4}, we list the various EOSs,  their particle compositions and properties of non-rotating neutron stars like maximum gravitational mass ($M_{G}$), radius $R_{max}$,  ($R_{1.4}$) and dimensionless tidal deformability $\Lambda_{1.4}$ of canonical mass. The properties like mass and radius for the neutron star are calculated by integrating the Tolman-Oppenheimer-Volkoff (TOV) equations   \citep{Weinberg1972}. TOV equations are solved for various EOSs consisting of nucleonic  and nucleonic with quark matter. The composition at any density is so determined that the charge neutrality and beta equilibrium conditions hold good.  A set of EOSs used in the present work has been displayed in Table \ref{tab:table4} where DOPS1, DOPS2 , and  DOPS3 are pure hadronic (nucleonic) EOSs computed with the newly generated DOPSs parameterizations. The EOSs namely  DOPS1Q, DOPS2Q, and DOPS3Q composed of nucleons and quarks in beta equilibrium with the NJL coexisting phase are also presented. For the sake of comparison, the results for EOSs calculated with NL3, FSUGarnet, IOPB-1 and Big Apple parameters are also presented.
\begin{figure}
\includegraphics[trim=0 0 0 0,clip,scale=0.37]{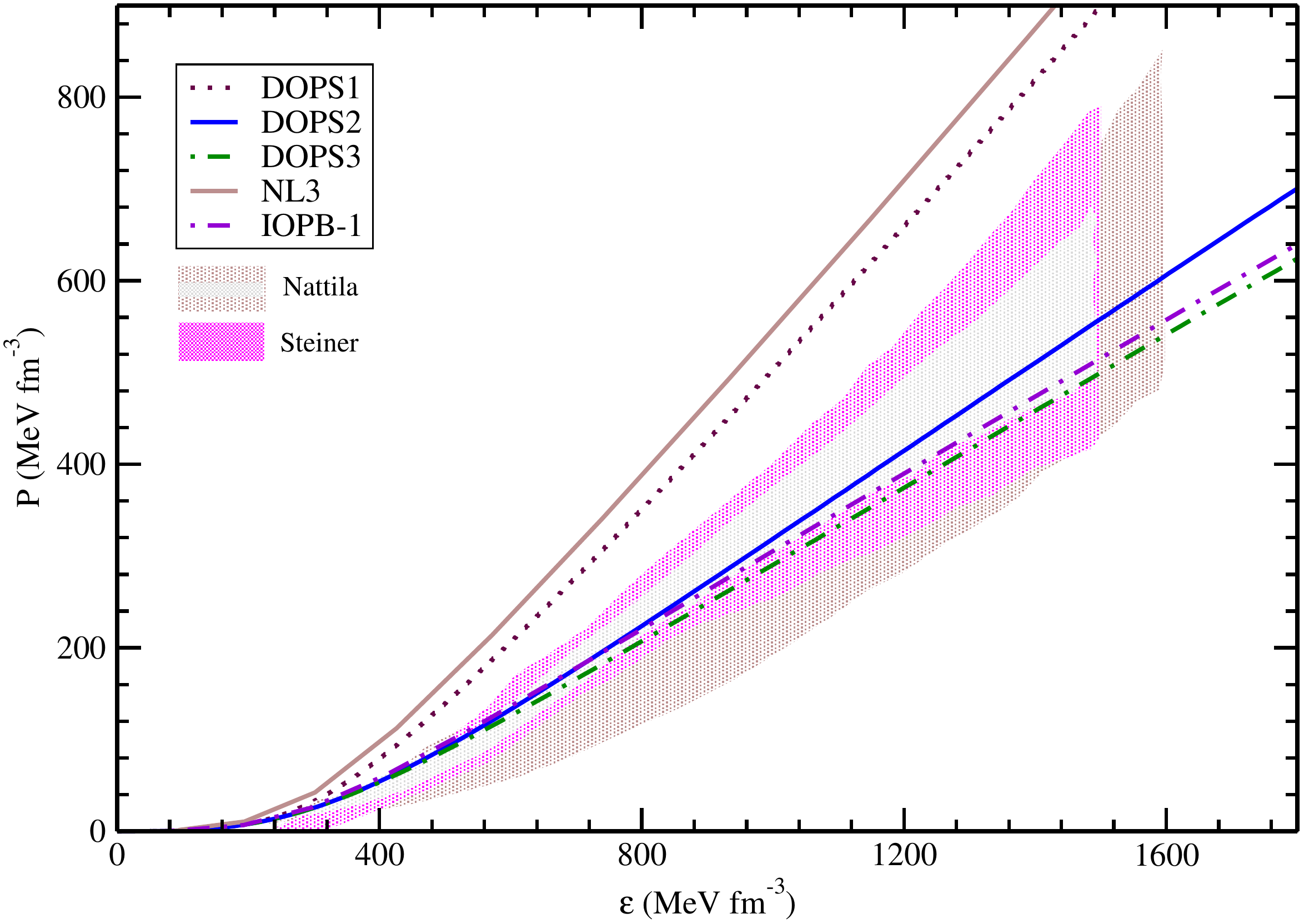}
\caption{\label{eos_nuclear} (Color online) Variation of pressure with energy density for EOSs calculated with DOPS1, DOPS2, DOPS3, NL3 and IOPB-1 parameterizations. The shaded region (magenta) represents the observational constrains from Ref.   \citep{Steiner2010} and the regions (orange and cyan) denote the EOS of cold dense matter with 95 $\%$ confidence limit   \citep{Nattila2016}. }
\end{figure}
In Fig. (\ref{eos_nuclear}), we display the  variation of pressure with energy density for various EOSs composed of pure nucleonic matter. The results are also compared with NL3 and IOPB-1 parameter sets. The shaded region (magenta color) represents the observational constraints from the ref.   \citep{Steiner2010} and regions (brown and grey) denote the EOS of cold dense matter with 95 $\%$ confidence limit   \citep{Nattila2016}. It is clear that EOS computed with DOPS1 parameter set is very stiff like NL3 and is  required to account for supermassive neutron star as predicted by GW190814 event and is ruled out by the above observational constraints. The EOSs calculated with  DOPS2, DOPS3 , and IOPB-1 are relatively softer and are in well agreement with the observational constraints as shown in  Fig. (\ref{eos_nuclear}).\\
\begin{figure}
\includegraphics[trim=0 0 0 0,clip,scale=0.36]{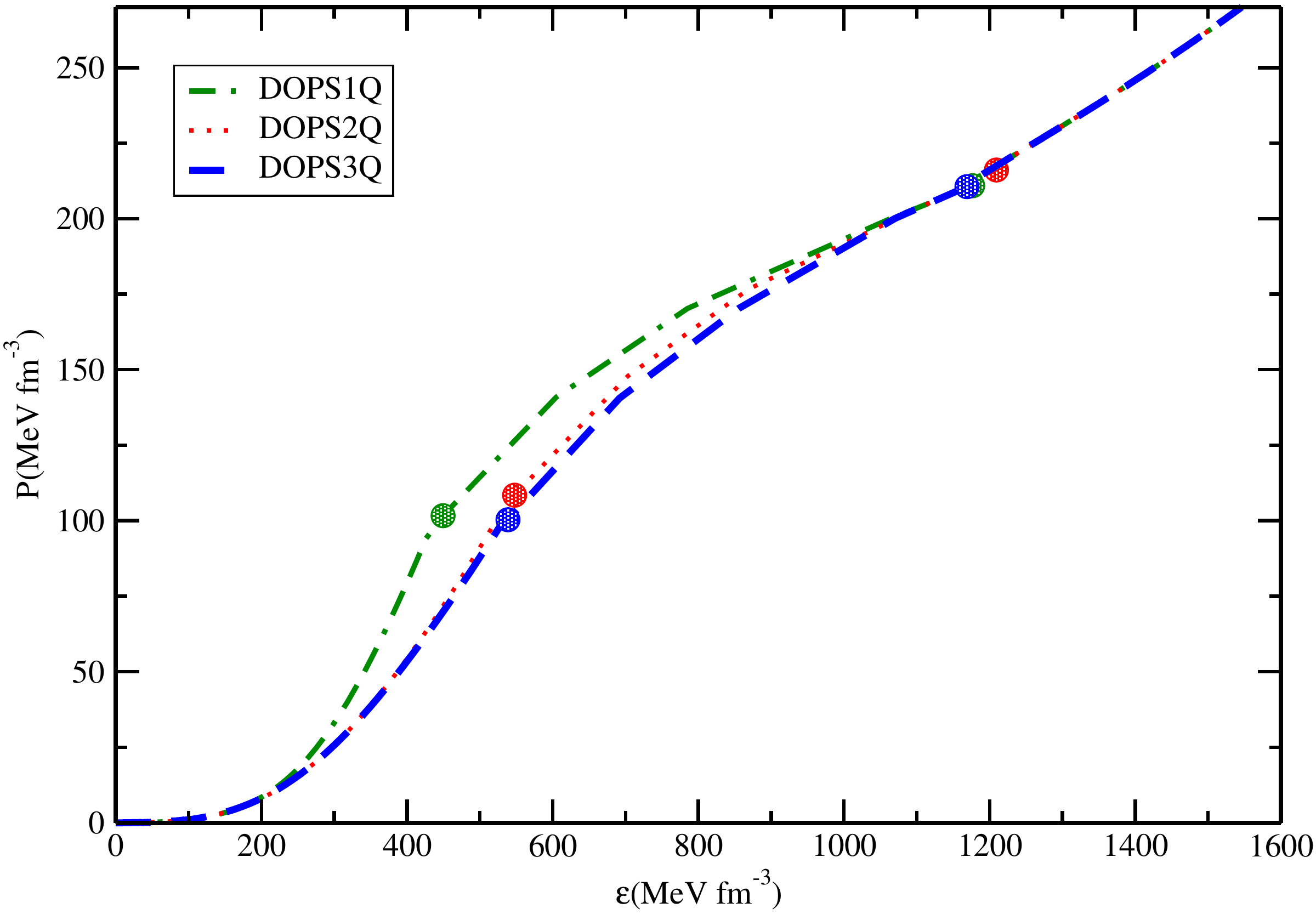}
\caption{\label{eos_mix} (Color online)  Variation of pressure with energy density for hybrid EOSs DOPS1Q, DOPSQ and DOPSQ3 composed of pure nucleonic matter,  quark matter and NJL coexisting phase.  The solid circles represent the boundary of coexisting phase comprised of nucleons and quarks.}
\end{figure}
In Fig. (\ref{eos_mix}), we plot the results for hybrid EOSs composed of nucleons and quark matter with a coexisting phase in $\beta$-equilibrium. The nucleonic part of the EOS is calculated by using DOPSs parameterizations and the pure quark phase is described with three flavor NJL model as discussed in Section II. For the coexisting phase of hadronic matter and quark matter, the Glendenning  construction method has been employed along with the global charge neutrality condition. The solid circles represent the boundary of coexisting phase region which consists of nucleons and quarks.
The coexisting phase region lies in the density ranging from 2.88$\rho_{0}$ - 6.5$\rho_{0}$,  3.49$\rho_{0}$- 6.71$\rho_{0}$ and 3.43$\rho_{0}$- 6.57$\rho_{0}$   for DOPS1, DOPS2 and DOPS3 respectively. The coexisting phase region for the DOPS1 region is large as compared to  others. As DOPS1 parameter set produces a stiff EOS and thus the coexisting phase region lies in the higher pressure region. \\
\begin{figure}
\includegraphics[trim=0 0 0 0,clip,scale=0.46]{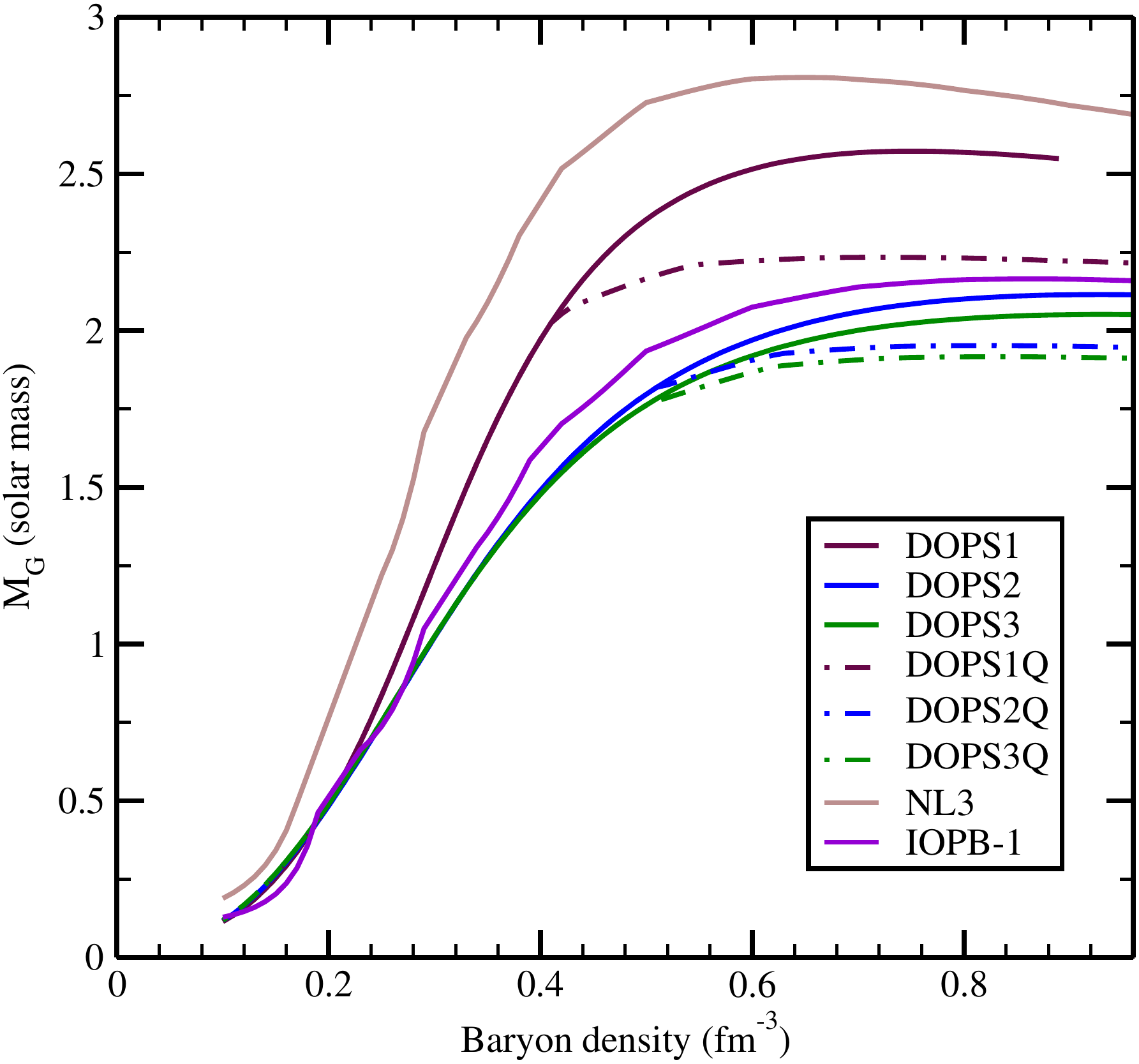}
\caption{\label{md} (Color online)  Gravitational mass ($M_{G}$) for the non-rotating  compact stars as a function of baryon density  for various EOSs.}
\end{figure}
In Fig. (\ref{md}), we plot the gravitational mass ($M_{G}$) of the neutron star as a function of the baryon  density  for various EOSs considered in the present work. It is evident from the figure  that gravitational mass increases with the increase in baryon  density to obtain its maximum value. It is quite obvious that the maximum gravitational mass is corresponding to the stiffest EOS and goes on decreasing as the EOS becomes softer.
\begin{figure}[h]
\includegraphics[trim=0 0 0 0,clip,scale=0.5]{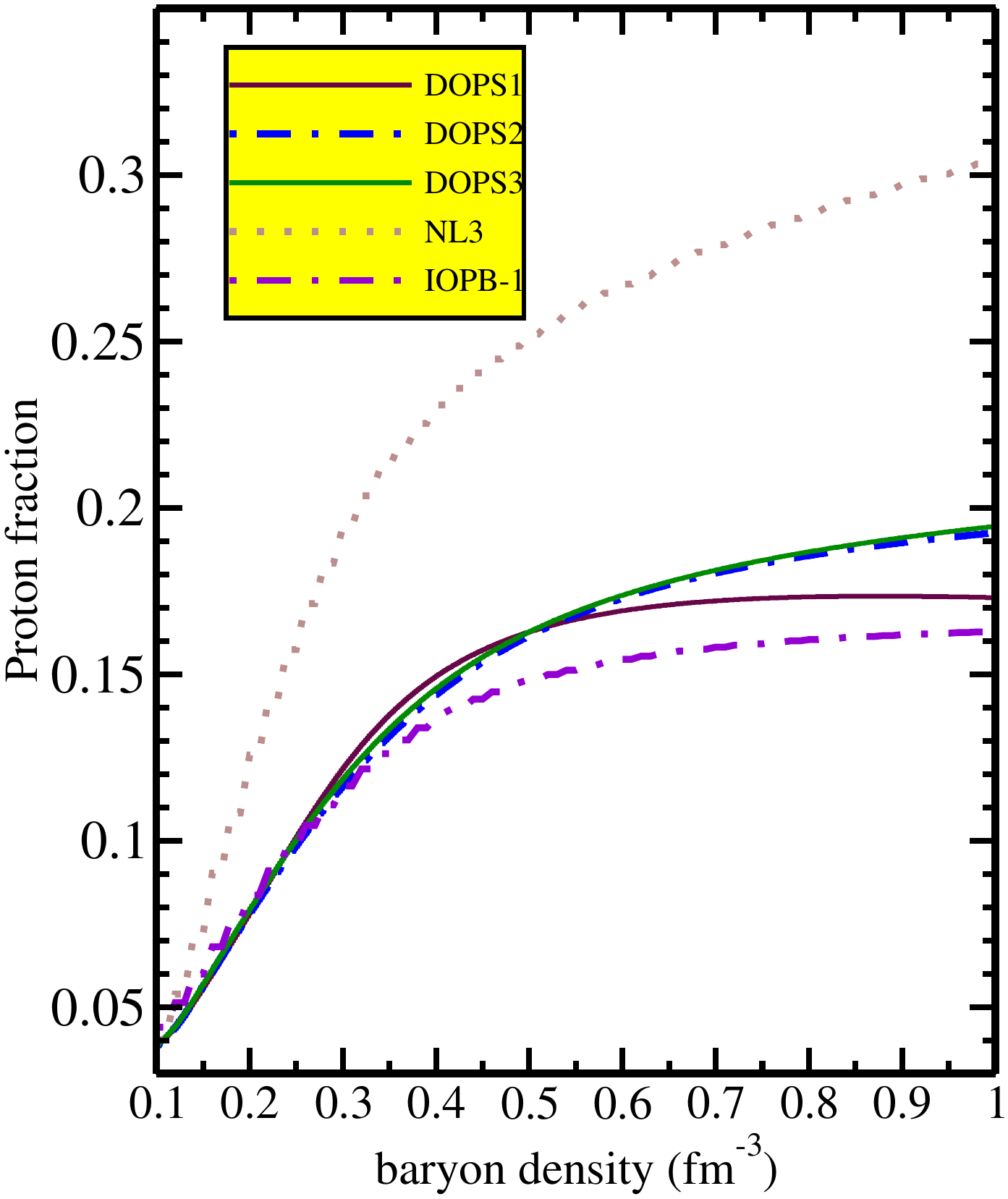}
\caption{\label{fraction} (Color online)  Proton fraction plotted against the baryon density.}
\end{figure}
In Fig. (\ref{fraction}), we have also shown the proton fraction as a function of baryon density for various EOSs.
\\
\begin{figure}
\includegraphics[trim=0 0 0 0,clip,scale=1.3]{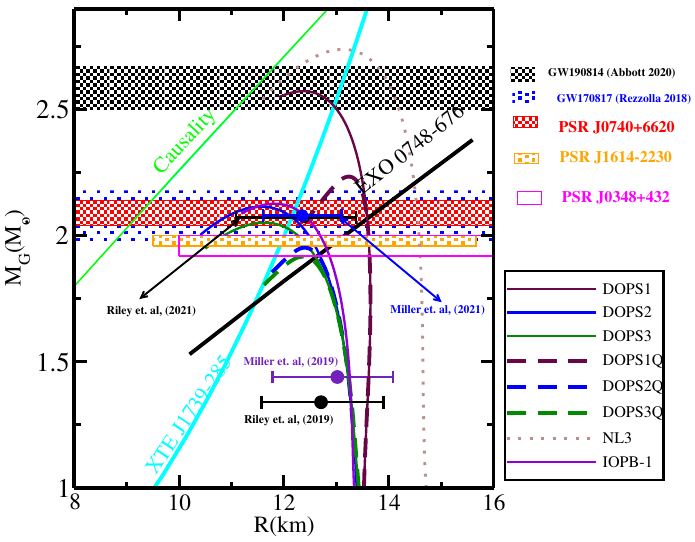}
\caption{\label{mr} (Color online) Mass-Radius profile for pure nucleonic and hybrid non-rotating neutron star for various EOSs. The recently observed constraints on  mass and radius  measurements from recently observed astrophysical events PSR J0740+6620   \citep{Riley2021,Miller2021,Riley2019,Miller2019}, PSR J1614+2230   \citep{Demorest2010}, PSR J0348+432   \citep{Antoniadis2013}, GW190814   \citep{Abbott2020} and GW170817   \citep{Rezzolla2018}  are also depicted. The region excluded by causality (solid green line), rotation constraint of neutron star XTE J1739-285 solid (solid cyan line) and limits on Mass-Radius of compact star from Ozel's analysis of EXO 0748-676 are also shown.} 
\end{figure}
 \begin{figure}
\includegraphics[trim=0 0 0 0,clip,scale=0.5]{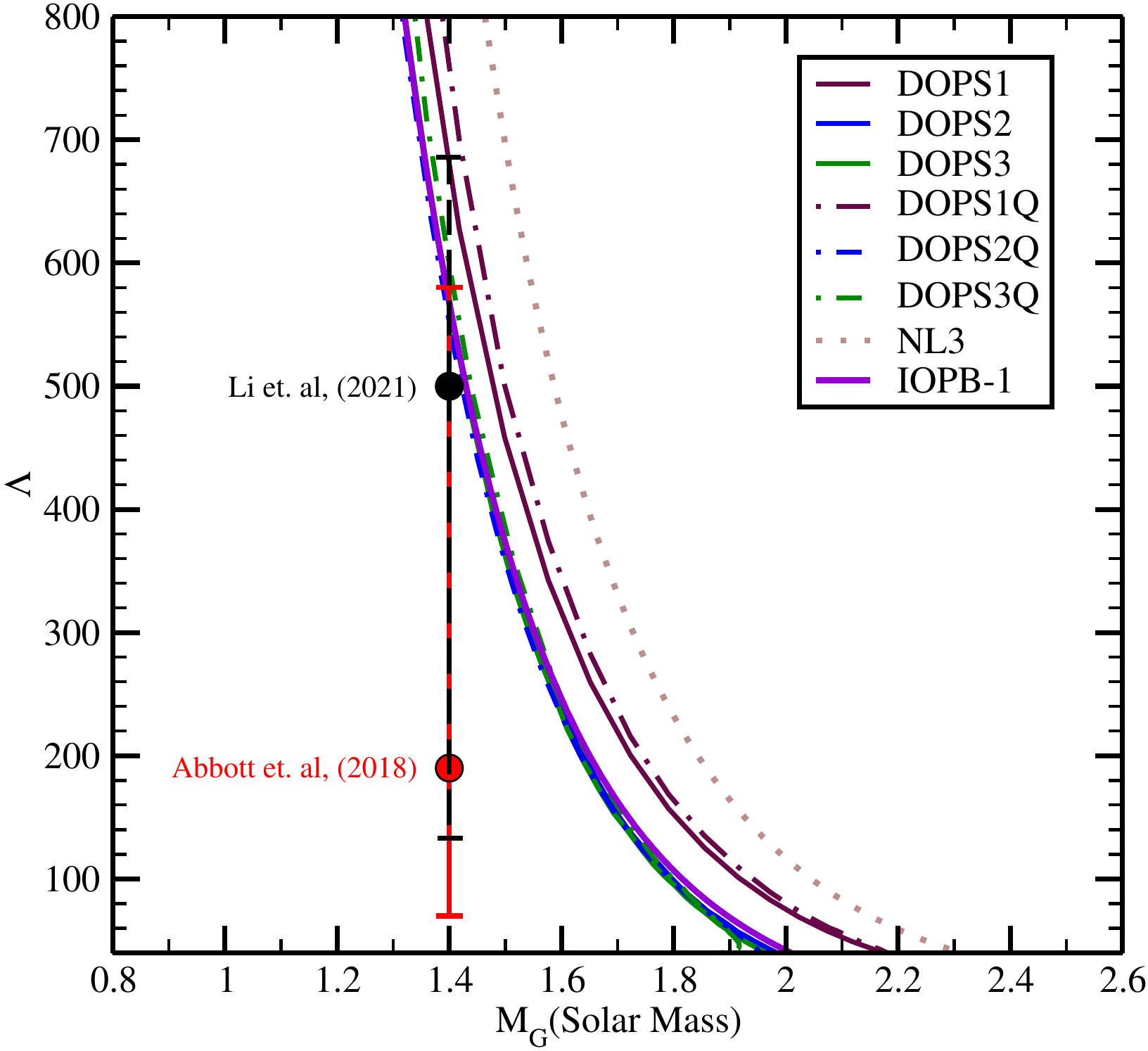}
\caption{\label{tidal} (Color online) Variation of  dimensionless tidal deformability ($\Lambda$)
 with respect
to gravitational mass of non-rotating compact stars for various EOSs. The recent constraints on $\Lambda_{1.4}$ from Ref.   \citep{Abbott2018,Li2021} are also depicted }
\end{figure}
Fig. (\ref{mr}), presents our results for the gravitational mass of non-rotating neutron star and its radius for DOPSs parameterizations. The results are also displayed for NL3 and IOPB-1 parameter sets. The maximum mass of non-rotating neutron star obtained for EOS calculated with the DOPS1 parameter set is found to be 2.57$M_{\odot}$  with  a radius of 12.36 Km. This maximum mass obtained with the DOPS1 parameter satisfies the  constraint from GW190814 event which gives the mass range between 2.50 - 2.67 $M_{\odot}$   \citep{Abbott2020} indicating that the secondary component might be the heaviest neutron star composed of nucleonic matter. This parameter set also satisfies the recently measured radius of PSR J0740+6620 with  = $12.39^{+1.30}_{-0.98}$Km  radius by NICER   \citep{Riley2021}.
 DOPS2 and DOPS3 sets produce non-rotating neutron stars of maximum mass 2.05$M_{\odot}$ and 2.12$M_{\odot}$ with  the radius of 11.56 Km and 11.61 Km respectively. The radii of canonical mass $R_{1.4}$ for DOPS2 and DOPS3 are calculated to be 13.24 Km and 13.25 Km respectively. The DOPS2 and DOPS3 parameter sets satisfy the mass constraints from GW170817   \citep{Rezzolla2018},  PSR 0740+6620,  NICER   \citep{Miller2021,Riley2021} and  radius constraints from  NICER   \citep{Riley2021,Riley2019,Miller2019,Annala2018} and is also very close to  the upper limit of the radius constraint   \citep{Miller2021}.
 The hybrid EOSs namely DOPS1Q, DOPS2Q , and DOPS3Q produce the hybrid  star with a maximum mass of 2.23, 1.95 , and 1.91 M$_{\odot}$  respectively. The phase transition from hadron to quark matter lowers the maximum mass as EOS becomes softer and , accordingly the  maximum mass reduces from 2.57 to 2.23 $M_{\odot}$, 2.12 to 1.95 $M_{\odot}$ and 2.05 to 1.91 $M_{\odot}$ corresponding to DOPS1, DOPS2 , and DOPS3 parameterizations respectively. The reduction in the maximum mass is found to be     more  in the stiffest EOS i.e. DOPS1. The hybrid  EOS  DOPS1Q satisfies the mass constraint of  MSP 0740+6620   \citep{Cromartie2020}. The  DOPS2Q and DOPS3Q EOSs satisfy the mass constraint from PSR J1614-2230   \citep{Demorest2010}. These hybrid EOSs satisfy the radius constraint from NICER   \citep{Riley2021,Annala2018}. The various observational constraints on maximum mass and radius  measurements from recently observed astrophysical events PSR J0740+6620   \citep{Riley2021,Miller2021}, PSR J1614+2230   \citep{Demorest2010}, PSR J0348+432   \citep{Antoniadis2013} and GW190814   \citep{Abbott2020} are also shown in Fig. (\ref{mr}).  Similar results for NL3 and IOPB-1 parameter sets are also displayed.\\
 The dimensionless tidal deformability ($\Lambda_{1.4}$) obtained by employing  EOSs namely DOPS2, DOPS3, DOPS2Q and DOPS3Q considered in the present work lie in the range 527 - 563 as one can see in Table \ref{tab:table4}. These values satisfy the constraint on $\Lambda_{1.4}$ = $190^{+390}_{-120}$   \citep{Abbott2018,Abbott2017}. The dimensionless tidal deformability ($\Lambda_{1.4}$) for EOSs DOPS1 and  DOPS1Q  are 627 and 637 respectively and is consistent with the constraints on dimensionless tidal deformability obtained using Bayesian analysics $\Lambda_{1.4}$ = $500^{+186}_{-367}$    \citep{Li2021}. 
 
 In Fig. (\ref{tidal}), we plotted  $\Lambda$  as a function of gravitational mass. It is obvious from the Fig. (\ref{tidal}) that $\Lambda$ decreases with the increase in gravitational mass of compact stars and reduces to very small value at the maximum mass. The recent observational limits on  $\Lambda_{1.4}$   \citep{Li2021,Abbott2018} are also displayed in Fig.  (\ref{tidal}).
\section{Summary} \label{s}
Theoretical studies of dense matter have considerable uncertainty in the
high density behavior of the EOSs largely because of the poorly constrained
many-body interactions. The theoretical study of the structure of neutron stars
is crucial if new observations of masses and radii lead to effective constraints on the EOSs of dense matter.
GW170817, GW190814 , and PSR J0740
+ 6620 etc. are some of those recent astrophysical observations (constraints) to
which our study is devoted. This study can further prompt the
theoretical/experimental(astrophysical observations) investigations.
The present study demonstrates that the contributions of self and mixed interactions of $\sigma$, $\omega$ , and $\rho$ mesons up to quartic order are important to vary the high density behavior of EOS and favor the naturalness behavior of parameters.\\
We have studied  the properties of finite nuclei, infinite nuclear matter , and compact stars with newly generated parameter sets DOPS1, DOPS2 , and DOPS3 of field theoretical relativistic mean field (RMF) models which include all the possible self and mixed interactions between the scaler-isoscalar ($\sigma$), vector-isoscalar ($\omega$) and vector-isovector ($\rho$) mesons up to quartic order. The generated parameter sets are in harmony with the finite and bulk nuclear matter properties. All these generated parameterizations fit equally well  the finite nuclear properties and  closely favor  the naturalness behavior   \citep{Furnstahl1997}. 
The mean absolute errors in the binding energy per nucleon calculated  with the DOPS1, DOPS2 , and DOPS3  parameterizations for the finite nuclei used in the fit are 0.027, 0.031, and 0.027 MeV respectively. Similarly,  the mean absolute error in the charge rms radii for  DOPS1, DOPS2, and DOPS3  parameterizations for the finite nuclei used in the fit are 0.019, 0.022 and 0.023 fm respectively. 
 The maximum mass of a non-rotating star with DOPS1 parameterization is found to be around 2.6 M$\odot$ for the pure hadronic matter which satisfies the recent GW190814 possible maximum mass constraint   \citep{Abbott2020}  indicating  that the secondary component of GW190814 could be the non-rotating neutron star consisting of pure nucleonic matter. This parameter set also satisfies the recently measured constraints on the radius  for PSR J0740+6620 with $12.39^{+1.30}_{-0.98}$Km by NICER   \citep{Riley2021}.
 DOPS2 and DOPS3 sets produce non-rotating neutron stars of maximum mass 2.05$M_{\odot}$ and 2.12$M_{\odot}$ with the  radius of 11.56 Km and 11.61 Km respectively. The radius of canonical mass $R_{1.4}$ for DOSPS2 and DOPS3 are calculated to be 13.24 Km and 13.25 Km respectively. The DOPS2 and DOPS3 parameter sets satisfy the mass constraints of   PSR 0740+6620,  NICER     \citep{Miller2021,Riley2021} and  radius constraints from  NICER   \citep{Riley2021,Riley2019,Miller2019,Annala2018} and is also very close the upper limit of the radius constraint   \citep{Miller2021}. EOSs computed with the DOPS2 and DOPS3 parameterizations also satisfy the X-Ray observational data by Steiner   \citep{Steiner2010} and the recent  observations of GW170817 maximum mass constraint of a stable non- rotating neutron star in the range 2.01 $\pm$ 0.04 - 2.16 $\pm$ 0.03 M$\odot$   \citep{Rezzolla2018}.  The hybrid EOSs obtained with the NJL model also satisfy astrophysical constraints on the maximum mass of a neutron star from PSR J1614-2230   \citep{Demorest2010}. The value of dimensionless tidal deformability ($\Lambda_{1.4}$) calculated by employing the EOSs DOPS2, DOPS3, DOPS2Q , and DOPS3Q is found to be in the range 527.3 - 563.2 which is consistent with  the waveform models analysis of GW170817   \citep{Abbott2018}. The value of ($\Lambda_{1.4}$) calculated with EOSs DOPS1 and DOPS1Q is 627.37 and 637 which is consistent with the constraint on $\Lambda_{1.4}$ obtained using Bayesian analysis   \citep{Li2021}.
\begin{acknowledgments}
Virender Thakur is highly thankful to  Himachal Pradesh University and DST-INSPIRE for  providing computational facility and financial assistance (Junior/Senior Research Fellowship).
\end{acknowledgments}
%

\end{document}